\documentclass[notitlepage,apj]{emulateapj}

\usepackage{wrapfig,lipsum,booktabs}
\usepackage{natbib}
\usepackage{amsmath, amssymb}
\usepackage{color}
\usepackage{afterpage}
\usepackage{verbatim}
\usepackage{graphicx}
\usepackage{enumitem}
\usepackage{float}
\usepackage{floatflt}

\newcommand{\simgt}{\,\hbox{\lower0.6ex\hbox{$\sim$}\llap{\raise0.6ex\hbox{$>$}}}\,}
\newcommand{\simlt}{\,\hbox{\lower0.6ex\hbox{$\sim$}\llap{\raise0.6ex\hbox{$<$}}}\,}

\begin{document}

\title{A Parameter Space Exploration of Galaxy Cluster Mergers II: Effects of Magnetic Fields}

\author{Bryan Brzycki}
\affiliation{Department of Astronomy, University of California, Berkeley, 501 Campbell Hall, Berkeley, CA 94720, USA}
\and
\author{John ZuHone}
\affiliation{Harvard-Smithsonian Center for Astrophysics, 60 Garden St., Cambridge, MA 02138, USA}

\begin{abstract}
The hot intracluster plasma in clusters of galaxies is weakly magnetized. Mergers between clusters produce gas compression and motions which can increase the magnetic field strength. In this work, we perform high-resolution non-radiative magnetohydrodynamics simulations of binary galaxy cluster mergers with magnetic fields, to examine the effects of these motions on the magnetic field configuration and strength, as well as the effect of the field on the gas itself. Our simulations sample a parameter space of initial mass ratios and impact parameters. During the first core passage of mergers, the magnetic energy increases via gas compression. After this, shear flows produce temporary, Mpc-scale, strong-field ``filament'' structures. Lastly, magnetic fields grow stronger by turbulence. Field amplification is most effective for low mass ratio mergers, but mergers with a large impact parameter can increase the magnetic energy more via shearing motions. The amplification of the magnetic field is most effective in between the first two core passages of each cluster merger. After the second core passage, the magnetic energy in this region gradually decreases. In general, the transfer of energy from gas motions to the magnetic field is not significant enough to have a substantial effect on gas mixing and the subsequent increase in entropy which occurs in cluster cores as a result. In the absence of radiative cooling, this results in an overall decrease of the magnetic field strength in cluster cores. In these regions, the final magnetic field is isotropic, while it can be significantly tangential at larger radii.
\end{abstract}

\section{Introduction} \label{sec:intro}

The largest gravitationally bounded objects in our universe are galaxy clusters. Most of the mass in galaxy clusters is comprised of dark matter (DM), which is believed to be largely collisionless \citep{zwi37, bah77}. Most of the baryonic material in galaxy clusters is comprised of a hot diffuse plasma called the intracluster medium \citep[hereafter ICM;][]{sar88}, which emits in X-rays. The last and smallest component of mass is that of the galaxies themselves, which have an effect on the cluster as a whole via feedback from stars and active galactic nuclei (AGN). Galaxy clusters allow us to study the interplay of these different forms of matter in a gravitationally bound system close to cosmological length scales. 

There is strong observational evidence that the ICM is weakly magnetized \citep{car02,fer08,fer12}. Synchrotron radio emission has been observed from sources such as radio halos and radio relics \citep{fer01,gov01,bur92,bac03,ven07,git07,gov09,gia11}. Furthermore, Faraday rotation of polarized emission can be measured for galaxy clusters in radio. Rotation measure (RM) studies place magnetic field strengths in clusters on the order of 0.1-10 $\mu$G, going up to tens of $\mu$G in cluster cool cores \citep{per91,tay93,fer95,fer99,tay02,tay06,tay07,bon10}. Such field strengths imply that the magnetic field itself is dynamically weak; this is typically parameterized using the plasma parameter $\beta = p_{\rm th}/p_{\rm B} \sim 100-1000$, where $p_{\rm th}$ and $p_{\rm B}$ are the thermal and magnetic pressures, respectively. RM maps have been developed for some clusters, from which it appears that the coherence length of cluster magnetic fields is on the order of 10 kpc or less. Studies have also used RM maps to infer the cluster magnetic field power spectrum, which indicate that the magnetic field power spectrum is similar to a Kolmogorov type ($P_B(k) \propto k^{-5/3}$, where $k$ is the wavenumber), depending on the assumed value for the coherence length of the field fluctuations \citep{vogt03,vogt05,mur04,gov06,gov09,gui08}. Despite the dynamical weakness of this field, it has important effects on the microphysical properties of the cluster plasma. The Larmor radii of electrons and ions ($\rho_{\rm L} \sim$~npc) are many orders of magnitude smaller than their mean free paths ($\lambda_{\rm mfp} \sim$~kpc), with the result that momentum and heat fluxes from the dissipative processes of viscosity and thermal conduction are highly anisotropic \citep{bra65, nar01, rob16}. Finally, cosmic-ray electrons radiate in the radio band via synchrotron emission in radio relics, radio halos, and radio mini-halos \citep{brunetti2014}.

Merging between galaxy clusters is responsible for forming new clusters and changing the state of both baryonic and dark matter within clusters. As many observed galaxy clusters show evidence of current or recent merging, it is important to understand the internal physics and observable properties of cluster mergers. Mergers compress the gas and generate shocks, cold fronts, and turbulent motions in the ICM. Because the magnetic field is effectively ``frozen-in'' the plasma, this compression and gas motions amplify and/or stretch magnetic field lines and increase the energy in the magnetic field. A number of recent studies have used magnetohydrodynamic (MHD) simulations in the cosmological context to investigate the mechanisms and efficiency by which gas compression, bulk flows, and turbulent gas motion can amplify magnetic fields in clusters \citep[e.g.][]{dol99,dolag2002,dolag2005,xu2009,xu2010,xu2011,vazza2014,marinacci2015,egan2016,vazza2018,domfern19}. For an extensive review of processes which amplify the magnetic field in clusters, see also \citet{donnert2018}.

Such magnetic field amplification may have important observable effects. It is well-known that magnetic fields stretched parallel to cold fronts by shear flows can stabilize them against Kelvin-Helmholtz instabilities \citep[][hereafter ZML11]{zuh11b} and at least partially suppress thermal conduction across them \citep{zuh13a,zuh15}. If the magnetic field strength is increased enough, it may produce a ``plasma depletion layer'' with a high magnetic field and low gas density and X-ray emissivity (ZML11). Evidence for such layers has been tenatively observed in a few clusters, including Virgo \citep{wer16}, A520 \citep{wang16}, and A2142 \citep{wang18}. The regions of amplified magnetic field also coincide with the same regions of increased turbulence, enhancing radio halo and radio mini-halo emission \citep{donnert2013,zuh13b,marinacci2018}.

It is also possible that an increased magnetic field can suppress gas mixing. During a cluster merger, the gas from the two clusters mixes. For ``cool-core'' clusters (characterized by temperature inversions, high gas densities, and low central entropies and cooling times), such mixing can increase the entropy of the core gas substantially \citep[][hereafter Z11]{mitchell2009,zuh11a}. If the magnetic pressure and tension are comparable to turbulent gas motions, they could resist this mixing and prevent this increase of the entropy of the gas (ZML11). 

In this work, we analyze MHD simulations of a parameter space of idealized binary cluster mergers with magnetic fields, and examine the properties of the gas and magnetic fields throughout the merger. In particular, we seek to determine the effect of various merger scenarios on the structure and strength of the magnetic field, and also determine its effects on the hot plasma. These simulations are complementary to cosmological simulations, since these controlled setups allow for a finer degree of control over cluster conditions. The simulations we use span a range of mass ratios and impact parameters. Though a number of past studies have included the effect of magnetic fields on idealized mergers or cluster substructure simulations \citep[e.g.][]{roe99,asai04,tak08,donnert2013,suz13,lage2014,vij17a,vij17b}, ours includes the most expansive study of different merger scenarios to date. We will also compare these simulations to the otherwise identical ``unmagnetized'' simulations from Z11, which do not contain magnetic fields.

This paper is organized as follows: In Section \ref{sec:methods}, we describe the methods, including the relevant physics, the simulation code, and the initial conditions. In Section \ref{sec:results}, we present our results. In Section \ref{sec:conclusions}, we summarize these results and their implications for physics within the cores of galaxy cluster mergers. We assume a $\Lambda$CDM cosmology with $h = 0.7$, $\Omega_m = 0.3$, and $\Omega_\Lambda = 0.7$.

\section{Methods}\label{sec:methods}

\subsection{Physics}\label{subsec:physics}

Our simulations solve the ideal MHD equations. Written in conservation form in Gaussian units, they are:
\begin{equation}
\frac{\partial \rho_g}{\partial t} + \nabla\cdot(\rho_g\mathbf{v})=0
\end{equation}
\begin{equation}
\frac{\partial(\rho_g \mathbf{v})}{\partial t} + \nabla\cdot\left(\rho_g\mathbf{v}\mathbf{v}-\frac{\mathbf{B}\mathbf{B}}{4\pi}\right)+\nabla p=\rho_g \mathbf{g} 
\end{equation}
\begin{equation}
\frac{\partial E}{\partial t} + \nabla\cdot\left[\mathbf{v}(E+p)-\frac{\mathbf{B}(\mathbf{v}\cdot\mathbf{B})}{4\pi}\right]=\rho_g\mathbf{g}\cdot\mathbf{v} 
\end{equation}
\begin{equation}\label{eqn:induction}
\frac{\partial B}{\partial t} + \nabla\cdot(\mathbf{v}\mathbf{B}-\mathbf{B}\mathbf{v})=0,
\end{equation}
where $\rho_g$ is the gas density, $\textbf{v}$ is the gas velocity, and $\textbf{B}$ is the magnetic field strength. The total energy $E$, total pressure $p$, and gravitational acceleration $\textbf{g}$ have the usual definitions:
\begin{equation}
p = p_{\text{th}}+\frac{B^2}{8\pi}
\end{equation}
\begin{equation}
E = \frac{1}{2}\rho v^2+\epsilon+\frac{B^2}{8\pi}
\end{equation}
\begin{equation}
\mathbf{g}=-\nabla\phi
\end{equation}
\begin{equation}
\nabla^2\phi=4\pi G(\rho_g+\rho_{\text{DM}})
\end{equation}
where $\epsilon$ is the gas internal energy per unit volume, and $\phi$ is the gravitational potential. We assume an ideal gas equation of state with $\gamma=5/3$.

\subsection{Simulation Code} \label{subsec:simulationcode}

We performed our simulations using FLASH, a parallel hydrodynamics/N-body astrophysical simulation code developed at the Center for Astrophysical Thermonuclear Flashes at the University of Chicago \citep{fry00,dub09}. FLASH employs adaptive mesh refinement (AMR), a method of partitioning a grid throughout the simulation box such that higher resolutions (smaller cell sizes) are only used where needed, such as in the cores of clusters and at the gas discontinuities formed in cluster mergers such as shocks and cold fronts.

FLASH solves the equations of magnetohydrodynamics using a directionally unsplit staggered mesh algorithm \citep[USM;][]{lee09}. The USM algorithm used in FLASH is based on a finite-volume, high-order Godunov scheme combined with a constrained transport method (CT), which guarantees that the evolved magnetic field satisfies the divergence-free condition \citep{eva88}. In our simulations, the order of the USM algorithm corresponds to the Piecewise-Parabolic Method (PPM) of \citet{col84}, which is ideally suited for capturing shocks and contact discontinuties (such as the cold fronts that appear in our simulations). FLASH also includes an $N$-body module which uses the particle-mesh method to solve for the forces on gravitating particles. The gravitational potential is computed using a multigrid solver included with FLASH \citep{ric08}.

\subsection{Initial Conditions} \label{subsec:initialconditions}

We carry out 9 simulations of idealized binary mergers of two spherically symmetric clusters in hydrostatic and virial equilibrium. Our initial galaxy clusters are initialized in the same way as in Z11, choosing initial conditions based on cosmological simulations and cluster observations, with the only difference that we now initialize the clusters with magnetic fields included. Since the simulations are otherwise identical to this previous set, we can compare each magnetized simulation to its corresponding unmagnetized version. 

These simulations have mass ratios $M_1/M_2$ of 1, 3, and 10, in which the primary cluster has an initial mass of $M_{200} = 6\times10^{14}~M_\odot$. From the mass, the rest of the parameters for each cluster can be derived using the scaling relations determined by \citet{vik09a}. In Table \ref{table1}, we list these parameters for each cluster, including some which are derived below.

In the rest of this section, we will describe the setup of our simulations, following closely the discussion from Z11. 

\subsubsection{Gas and Dark Matter}\label{subsec:gas_dm}

\begin{table*}[ht!]
\caption{Initial Cluster Parameters}
\begin{center} 
\begin{tabular}{ c c c c c c c c c } 
    \hline 
    Cluster & $M_{200}$ ($M_\odot$)$^a$ & $r_{200}$ (kpc)$^b$ & $c_{200}$$^c$ & $f_{g,500}$$^d$ & $T_X$ (keV)$^e$ & $S_0$ (keV cm$^2$)$^f$ & $S_1$ (keV cm$^2$)$^g$ & $N_p$$^h$ \\ \hline \hline
    C1 & $6\times 10^{14}$ & 1552.25 & 4.5 & 0.1056 & 4.97 & 9.62 & 192.40 & 5,000,000 \\ 
    C2 & $2\times 10^{14}$ & 1076.27 & 4.7 & 0.0879 & 2.42 & 5.08 & 101.60 & 1,684,119 \\ 
    C3 & $6\times 10^{13}$ & 720.49 & 5.1 & 0.0686 & 1.10 & 2.73 & 54.60 & 513,137 \\ 
    \hline
  \end{tabular} 
  \label{table1}
      \\$^a$Virial mass; $^b$Virial radius; $^c$Concentration parameter; $^d$Gas mass fraction; $^e$X-ray temperature; $^f$Core entropy; $^g$Scale entropy; $^h$Number of DM particles
  \end{center}
\end{table*}

\begin{table}
\caption {Initial Merger Parameters} 
\begin{center} 
\begin{tabular}{ c c c } 
    \hline 
    Simulation & $M_1/M_2$ & $b/r_{200}$ \\ \hline \hline
    MS1,\ S1 & 1 & 0 \\ 
    MS2,\ S2 & 1 & 0.3 \\ 
    MS3,\ S3 & 1 & 0.6 \\ 
    MS4,\ S4 & 3 & 0 \\ 
    MS5,\ S5 & 3 & 0.3 \\ 
    MS6,\ S6 & 3 & 0.6 \\ 
    MS7,\ S7 & 10 & 0 \\ 
    MS8,\ S8 & 10 & 0.3 \\ 
    MS9,\ S9 & 10 & 0.6 \\
    \hline
    
  \end{tabular} 
  \label{table2}
  \end{center}
\end{table}

For the total mass distribution, we use the Navarro-Frenk-White (NFW) density profile \citep{nav97}:
\begin{equation}
\rho_{\text{tot}}(r)=\frac{\rho_s}{r/r_s\left(1+r/r_s\right)^2},
\end{equation}
with
\begin{equation}
r_s = r_{200}/c_{200},
\end{equation}
\begin{equation}
\rho_s = \frac{200}{3}c_{200}^3\rho_{\text{crit}}\left[\log\left(1+r/r_s\right)-\frac{r/r_s}{1+r/r_s}\right]^{-1}.
\end{equation}

We carry the NFW profile out to the virial radius $r=r_{200}$, the radius at which the average density is 200 times the critical density of the universe $\rho_{\text{crit}}$. Then, for $r>r_{200}$, we employ an exponentially decreasing density prescription:
\begin{equation}
\rho_{\text{tot}}(r)=\frac{\rho_s}{c_{200}(1+c_{200})^2}\left(\frac{r}{r_{200}}\right)^{\kappa}\exp\left({-\frac{r-r_{200}}{0.1r_{200}}}\right),
\end{equation}
where $\kappa$ is a constant such that $\rho_{\text{tot}}$ and its first derivative are continuous at the boundary $r=r_{200}$. We employ this exponential cutoff, since the NFW mass profile does not converge as $r$ goes to infinity.

We take the gas to be in hydrostatic equilibrium in the DM-dominated potential well. A key observable quantity for the ICM is the gas entropy, defined as $S = k_BTn_e^{-2/3}$, where $k_BT$ is the gas temperature in keV and $n_e$ is the electron number density. The entropy profiles of galaxy clusters can be well-modeled by a ``baseline'' power-law profile combined with a constant floor value to represent the entropy of the core \citep{voi05,cav09}, which can be written as:
\begin{equation}
S(r)=S_0+S_1\left(\frac{r}{0.1r_{200}}\right)^{\alpha} \label{eqn:S_powerlaw}
\end{equation} where $S_0$ is the core entropy and $\alpha \sim 1.0-1.3$. We start off with small core entropies $S_0$ and set $\alpha=1.1$ in order to make our initial models consistent with relaxed, ``cool-core'' galaxy clusters.

Using the above definition of entropy, we can take the equation of hydrostatic equilibrium and derive an equivalent expression in terms of the gas entropy and temperature:
\begin{eqnarray}
{dP \over dr} &=& -\rho_g{d\phi \over dr} \\
{k_B \over {\mu}m_p}{d({\rho_g}T) \over dr} &=& -\rho_g{d\phi \over dr} \\
{k_B \over {\mu}m_p}{d \over dr}\left[{T{\left(T \over S\right)}^{3/2}}\right] &=& -\left({T \over S}\right)^{3/2}{d\phi \over dr}
\end{eqnarray}

We solve this equation using standard numerical integration methods and by imposing two conditions: the gas mass fraction $f_{g,500} = M_{\text{g}}(r_{500})/M_{\text{tot}}(r_{500})$ (see Table \ref{table1}) using the scaling relation from \citet{vik09a} and setting $T(r_{200})=\frac{1}{2} T_{200}$, where 
\begin{equation}
k_B T_{200} \equiv \frac{ G M_{200} \mu m_p}{2r_{200}}
\end{equation} is the ``virial temperature'' of the cluster \citep{poo06}. From the temperature and entropy, we determine the gas density profile. The DM density profile is then given by
\begin{equation}
\rho_{\text{DM}}(r)=\rho_{\text{tot}}(r)-\rho_g(r).
\end{equation}

After determining the initial radial profiles, we set up the distribution of positions and velocities for the DM particles, following the procedure outlined in \citet{kaz04}. For the positions, we uniformly sample a random deviate $u\in [0,1]$, and we invert the function $u=M_{\text{DM}}(r)/M_{\text{DM}}(r_{\text{cut}})$ to calculate the radius of each particle from the center of the DM halo, where $r_{\text{cut}}$ is the radius of the halo at which the gas density reaches the mean gas density of the universe, which we take to be the boundary of the halo, and is typically a few $r_{200}$. We choose to directly calculate the velocity distribution using the energy distribution function \citep{edd16}:
\begin{equation}
\mathcal{F} = \frac{1}{\sqrt{8} \pi^2}\left[\int_0^{\mathcal{E}} \frac{d^2\rho}{d\psi^2}\frac{d\psi}{\sqrt{\mathcal{E}-\psi}}+\frac{1}{\sqrt{\mathcal{E}}}\left(\frac{d\rho}{d\psi}\right)_{\psi=0}\right], 
\end{equation} where $\psi = -\phi$ is the relative potential and $\mathcal{E}=\psi - \frac{1}{2} v^2$ is the relative energy of the particle. Particle speeds are determined using this distribution function using the acceptance-rejection method. With the particle radii and speeds determined, we find the position and velocity vectors by choosing random unit vectors isotropically distributed in $\mathbb{R}^3$. The number of DM particles we use for each cluster is shown in Table \ref{table1}.

\subsubsection{Magnetic Fields}

The magnetic field of the cluster is set up after the manner of ZML11. This procedure is designed to produced a tangled magnetic field with a magnetic pressure roughly proportional to the thermal pressure, satisfying the condition $\nabla\cdot\mathbf{B}=0$. A Gaussian random magnetic field $\tilde{\bf B}({\bf k})$ is set up in $\bf{k}$-space on a uniform grid using independent normal random deviates for the real and imaginary components of the field. We adopt a dependence of the magnetic field amplitude $B(k)$ on the wavenumber $|\bf{k}|$ similar to (but not the same as) \citet{rus07} and \citet{rus10}:
\begin{equation}
B(k) \propto k^{-11/6}{\rm exp}[-(k/k_0)^2]{\rm exp}[-(k_1/k)^2]
\end{equation}
which corresponds to a Kolmogorov power spectrum with exponential cutoffs at scales of $k_0$ and $k_1$. The cutoff at high wavenumber $k_0 = 2\pi/\lambda_0$ is set to $\lambda_0$ = 1~kpc, though the finest cell size of $\Delta{x} \approx 7$~kpc effectively sets the smallest scale. The cutoff at low wavenumber (large length scale) $k_1 = 2\pi/\lambda_1$ corresponds to $\lambda_1 \approx$~500~kpc. This field is then Fourier transformed to yield ${\bf B}({\bf x})$, which is rescaled to have an average value of $\sqrt{8\pi{p}/\beta}$ to yield a field that has a pressure that scales with the gas pressure, i.e. to have a spatially uniform $\beta$ for the initial field. The value $\beta = 200$ is chosen to produce magnetic fields which agree with typical field measurements from Faraday rotation measurements \citep{bon10} and simulations \citep{dol99, dub08}. Recently, \citet{walker17,walker18} showed that simulations with an initial $\beta = 200$ from ZML11 provide the best match to conditions seen in the Perseus Cluster, further motivating our choice. 

\subsubsection{Merger Trajectories}

In each merger simulation, we set up two clusters centered within a cubical simulation box of $\sim$14.29~Mpc on a side. The boundary conditions of the simulation box are such that matter may flow into and out of the box. We find that mass loss through these boundaries is negligible and does not affect the evolution of our cluster mergers, which occur in the central $\sim$(6~Mpc)$^3$ region. 

The cluster centers are initialized in the $x-y$ coordinate plane at $z$ = 0, and the initial distance between them is given by the sum of their respective $r_{200}$. \citet{vit02} demonstrated from cosmological simulations that the average infall velocity for merging clusters is $v_{\rm in}(r_{\rm vir}) = 1.1V_c$, where $V_c = \sqrt{GM(r_{\rm vir})/r_{\rm vir}}$ is the circular velocity at the virial radius $r_{\rm vir}$ for the primary cluster. For all of our simulations, this is chosen as the initial relative velocity ($v_{\rm in} \approx 1200$ km/s). In addition to varying the mass ratio between the simulation, we also vary the impact parameter of each merger simulation between $b$ = 0 (head-on), $0.3r_{200}$, and $0.6r_{200}$. In Table \ref{table2}, we list all of the simulations, referring to the magnetized simulations with prefix ``MS'' and unmagnetized with prefix ``S''.

\section{Results} \label{sec:results}

\subsection{Slices in Density and Magnetic Field Strength} \label{subsec:slices}

\begin{figure*}
\begin{center}
\includegraphics[width=0.94\linewidth]{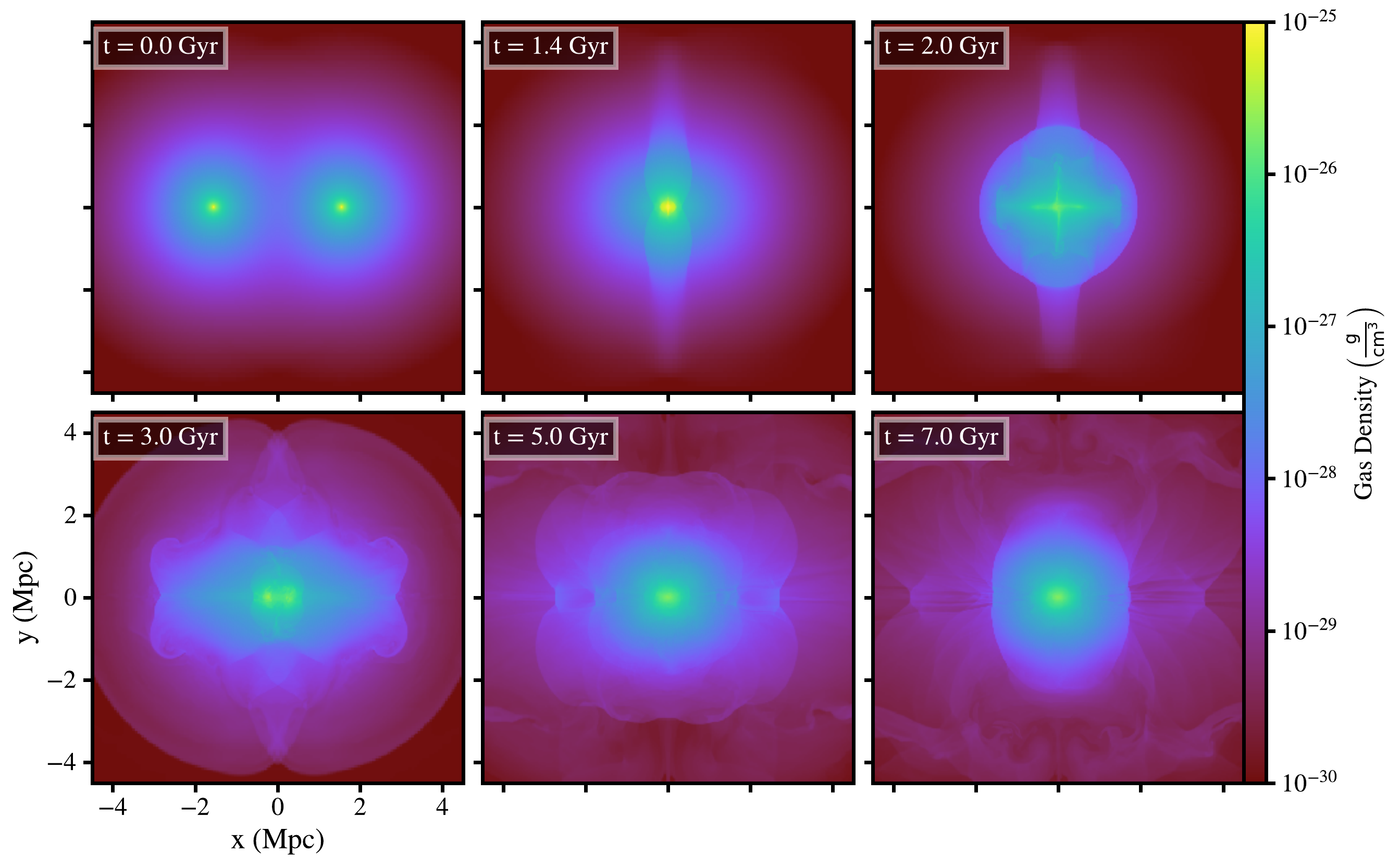}
\end{center}
\caption{Density slices through the collision axis for the MS1 simulation ($M_1/M_2 = 1$, $b = 0$). The epochs shown are: $t$ = 0, 1.4, 2.0, 3.0, 5.0, and 7.0 Gyr.\label{fig:1to1_b0_density}}
\end{figure*} 

\begin{figure*}
\begin{center}
\includegraphics[width=0.94\linewidth]{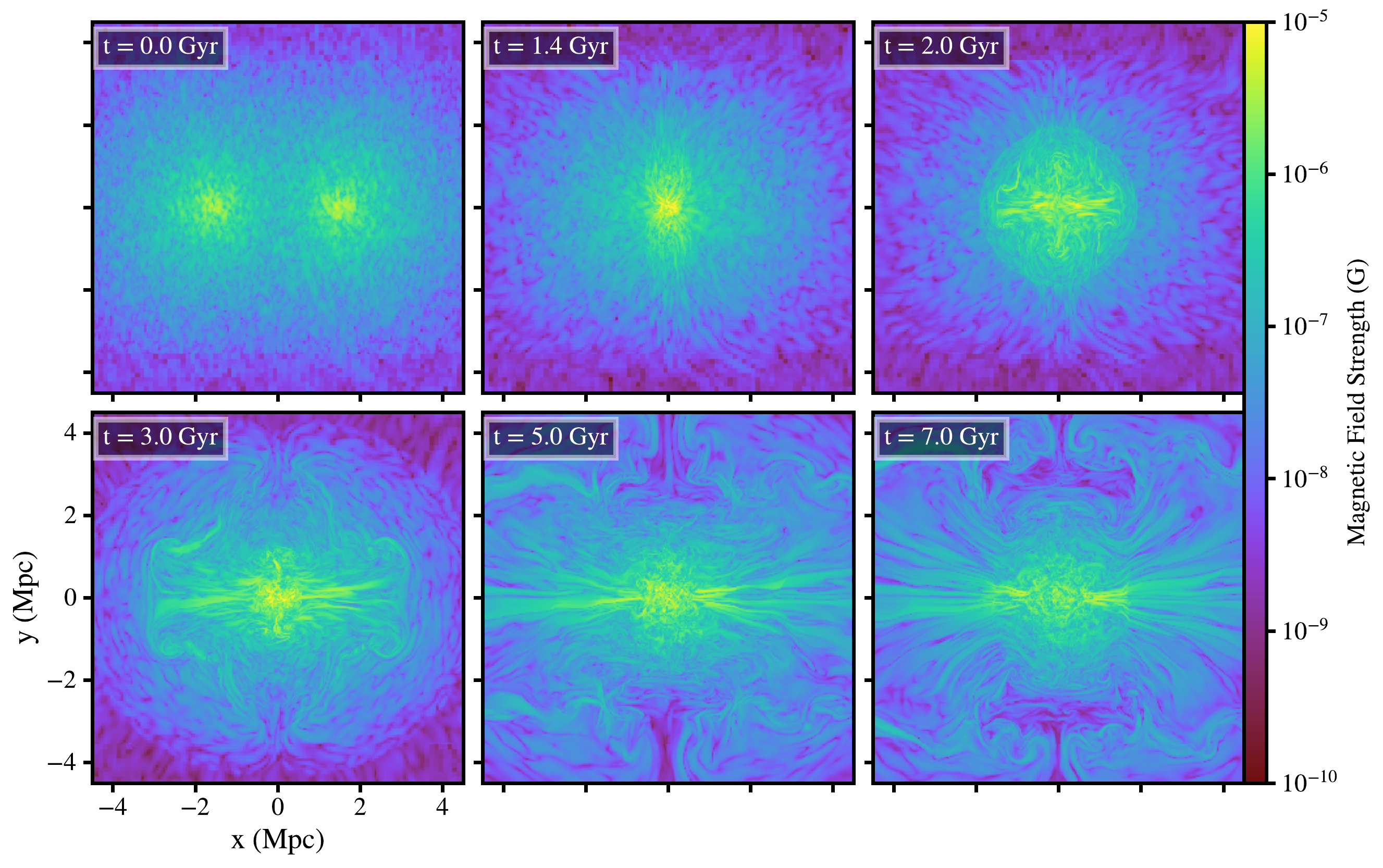}
\end{center}
\caption{Magnetic field strength slices through the collision axis for the MS1 simulation ($M_1/M_2 = 1$, $b = 0$). The epochs shown are: $t$ = 0, 1.4, 2.0, 3.0, 5.0, and 7.0 Gyr.\label{fig:1to1_b0_magfield}}
\end{figure*} 

\begin{figure*}
\begin{center}
\includegraphics[width=0.94\linewidth]{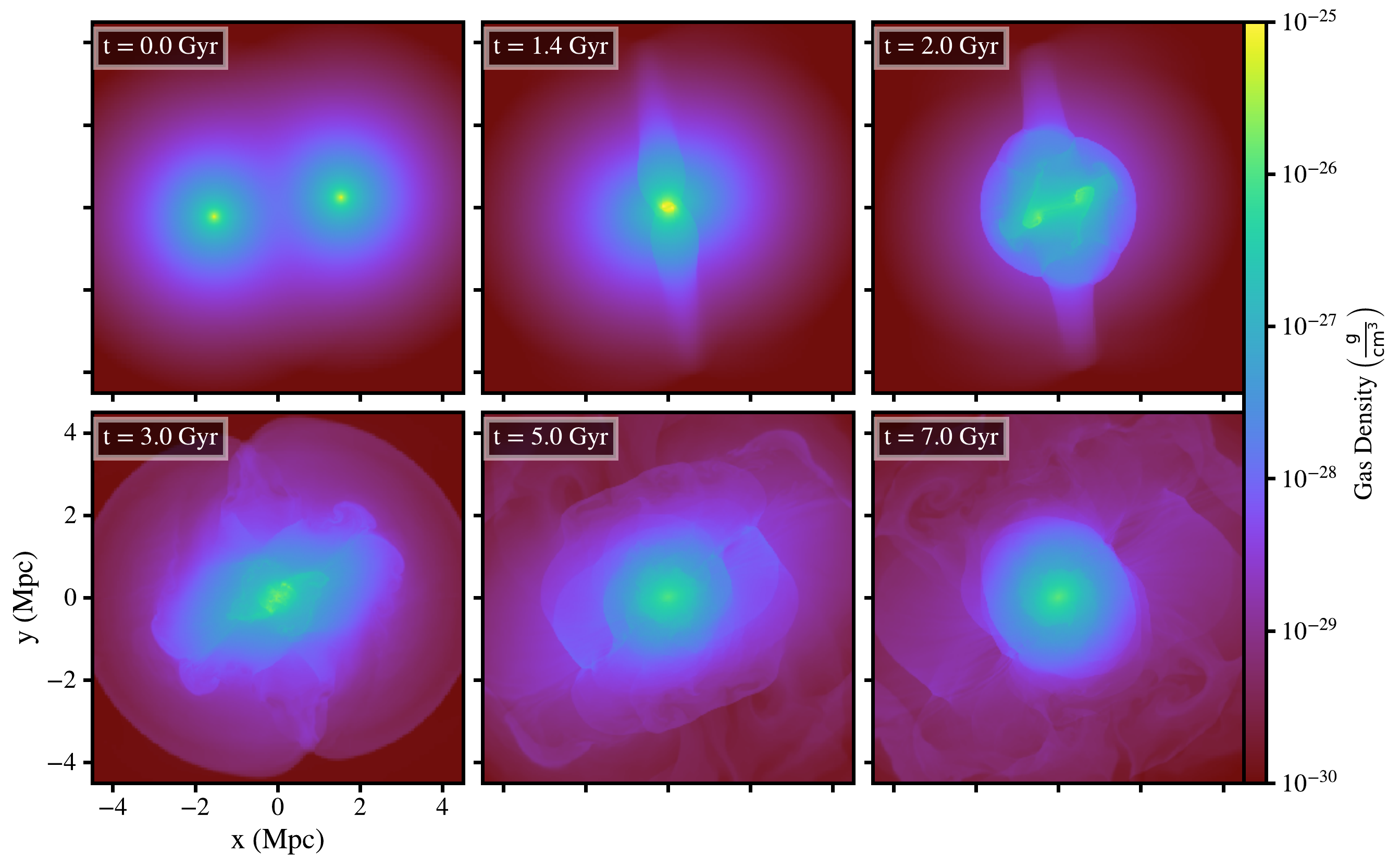}
\end{center}
\caption{Density slices through the collision axis for the MS2 simulation ($M_1/M_2 = 1$, $b = 0.3r_{200}$). The epochs shown are: $t$ = 0, 1.4, 2.0, 3.0, 5.0, and 7.0 Gyr.\label{fig:1to1_b0.5_density}}
\end{figure*} 

\begin{figure*}
\begin{center}
\includegraphics[width=0.94\linewidth]{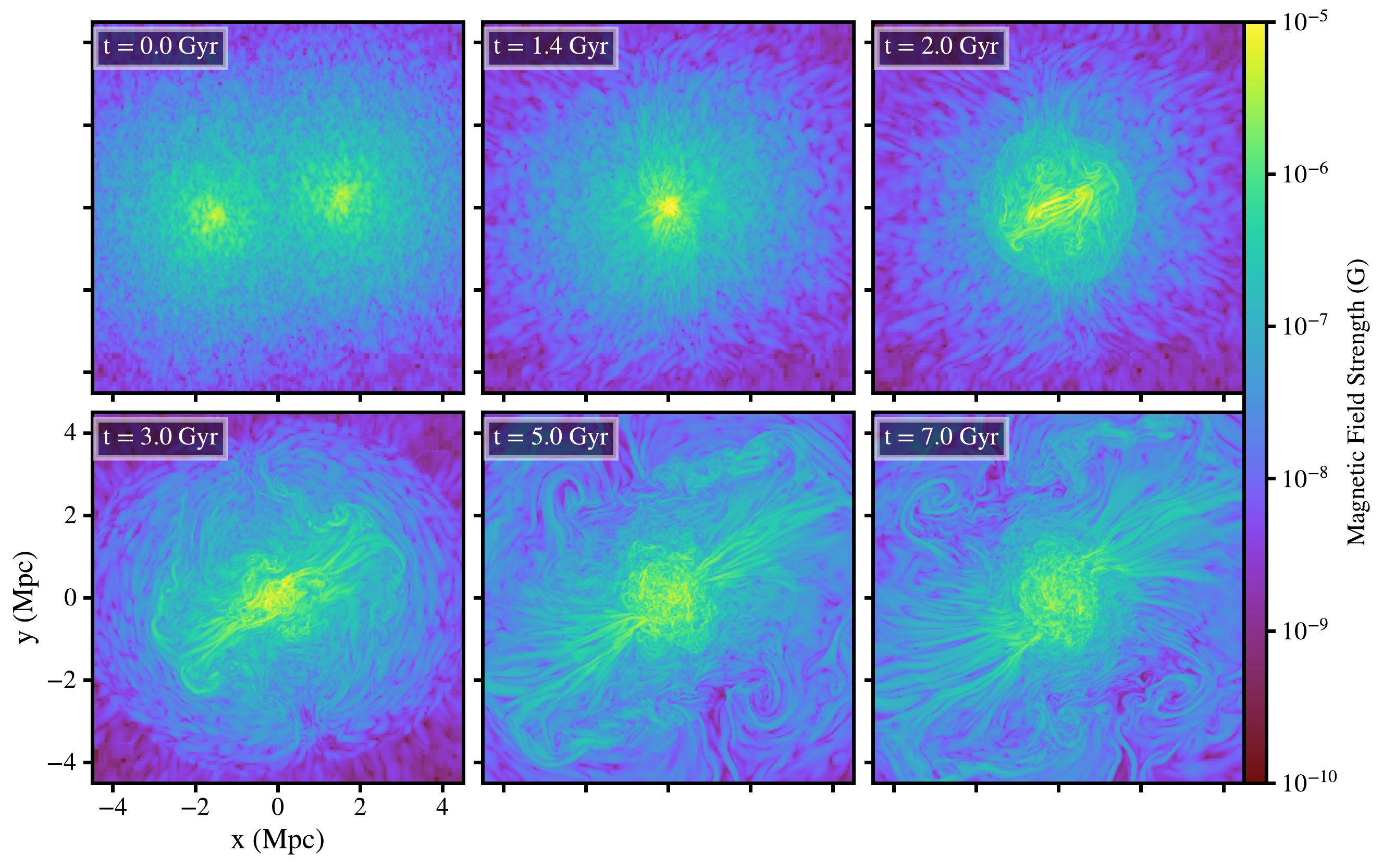}
\end{center}
\caption{Magnetic field strength slices through the collision axis for the MS2 simulation ($M_1/M_2 = 1$, $b = 0.3r_{200}$). The epochs shown are: $t$ = 0, 1.4, 2.0, 3.0, 5.0, and 7.0 Gyr.\label{fig:1to1_b0.5_magfield}}
\end{figure*} 

\begin{figure*}
\begin{center}
\includegraphics[width=0.94\linewidth]{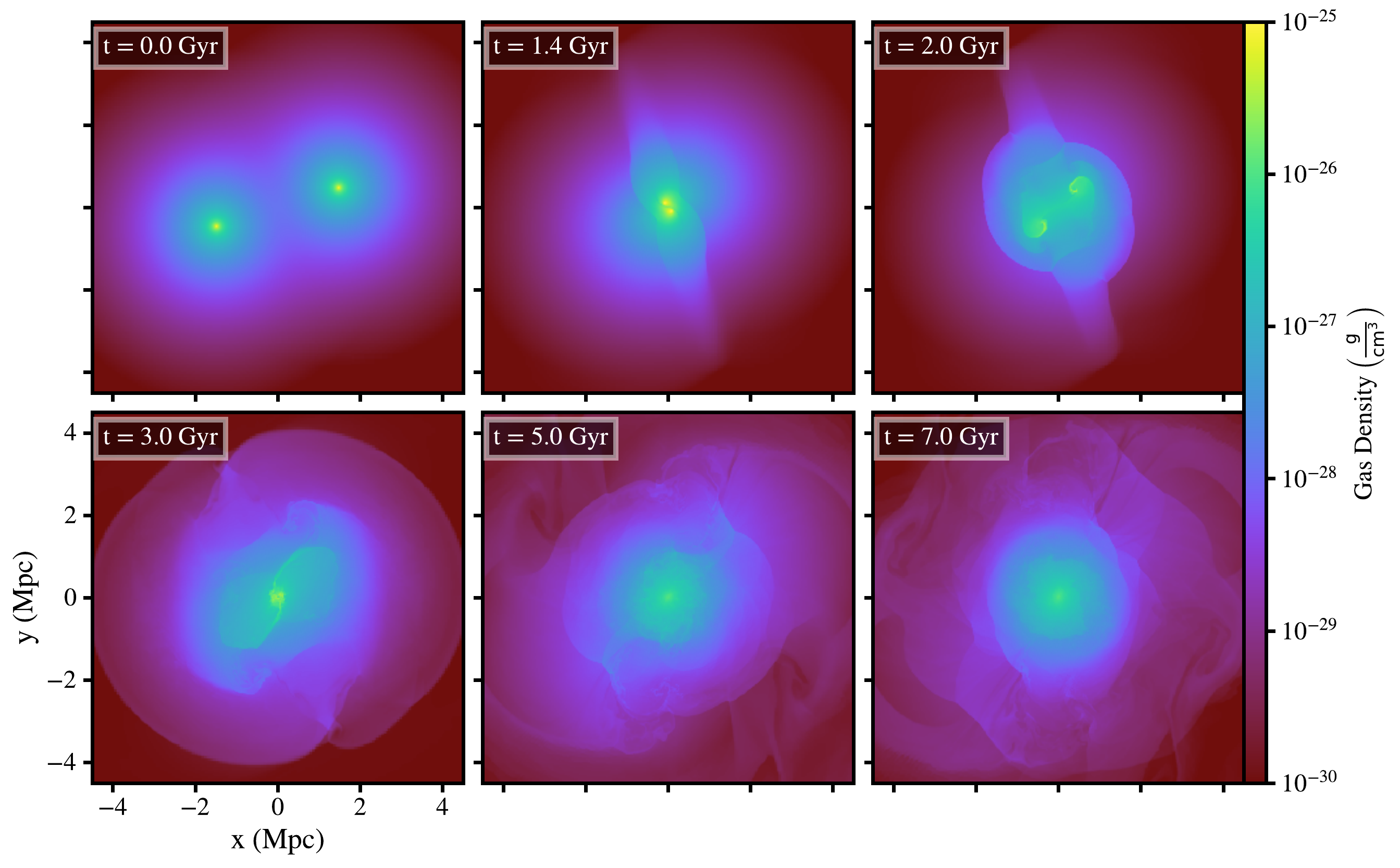}
\end{center}
\caption{Density slices through the collision axis for the MS3 simulation ($M_1/M_2 = 1$, $b = 0.6r_{200}$). The epochs shown are: $t$ = 0, 1.4, 2.0, 3.0, 5.0, and 7.0 Gyr.\label{fig:1to1_b1_density}}
\end{figure*} 

\begin{figure*}
\begin{center}
\includegraphics[width=0.94\linewidth]{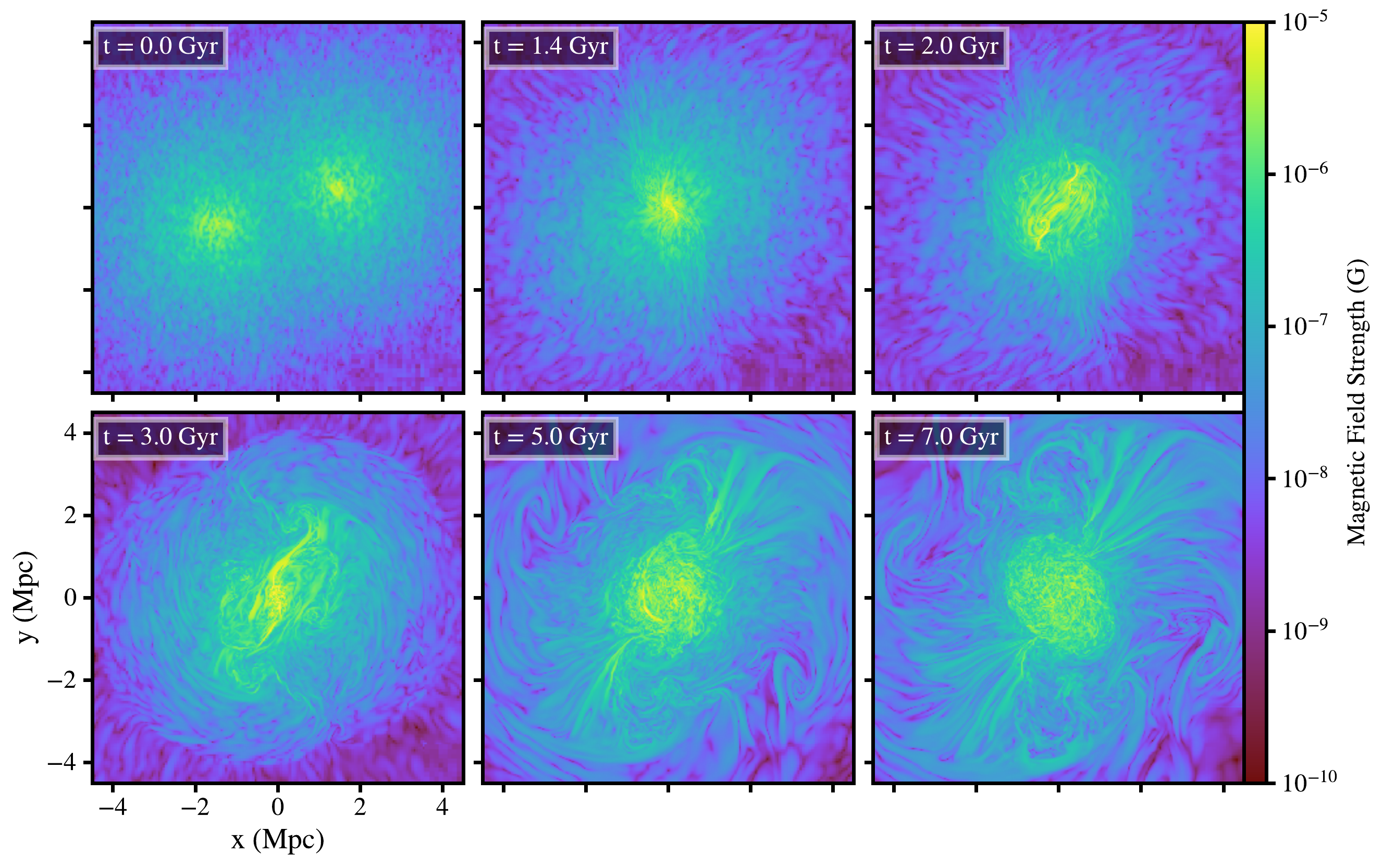}
\end{center}
\caption{Magnetic field strength slices through the collision axis for the MS3 simulation ($M_1/M_2 = 1$, $b = 0.6r_{200}$). The epochs shown are: $t$ = 0, 1.4, 2.0, 3.0, 5.0, and 7.0 Gyr.\label{fig:1to1_b1_magfield}}
\end{figure*} 

\begin{figure*}
\begin{center}
\includegraphics[width=0.94\linewidth]{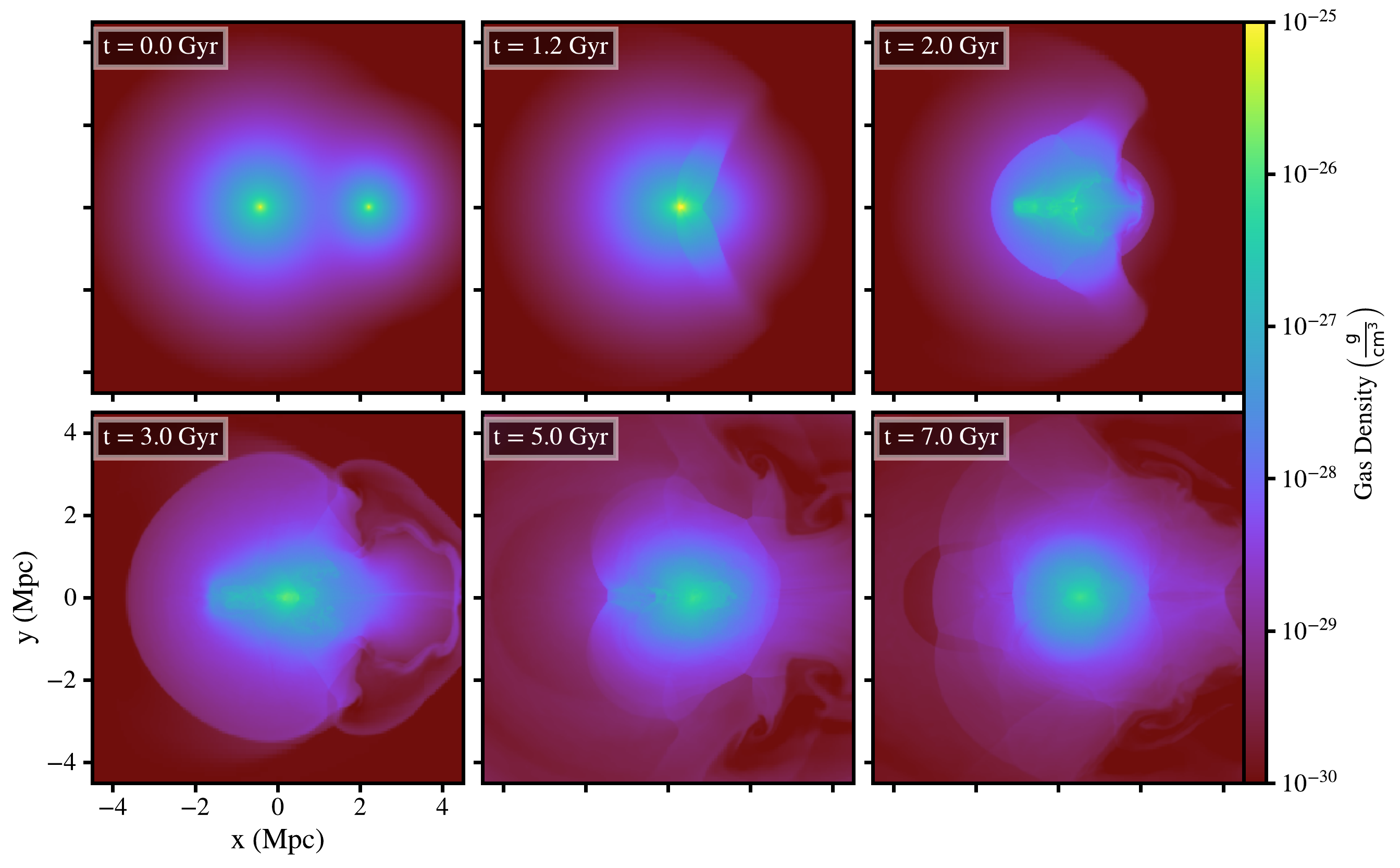}
\end{center}
\caption{Density slices through the collision axis for the MS4 simulation ($M_1/M_2 = 3$, $b = 0$). The epochs shown are: $t$ = 0, 1.2, 2.0, 3.0, 5.0, and 7.0 Gyr.\label{fig:1to3_b0_density}}
\end{figure*} 

\begin{figure*}
\begin{center}
\includegraphics[width=0.94\linewidth]{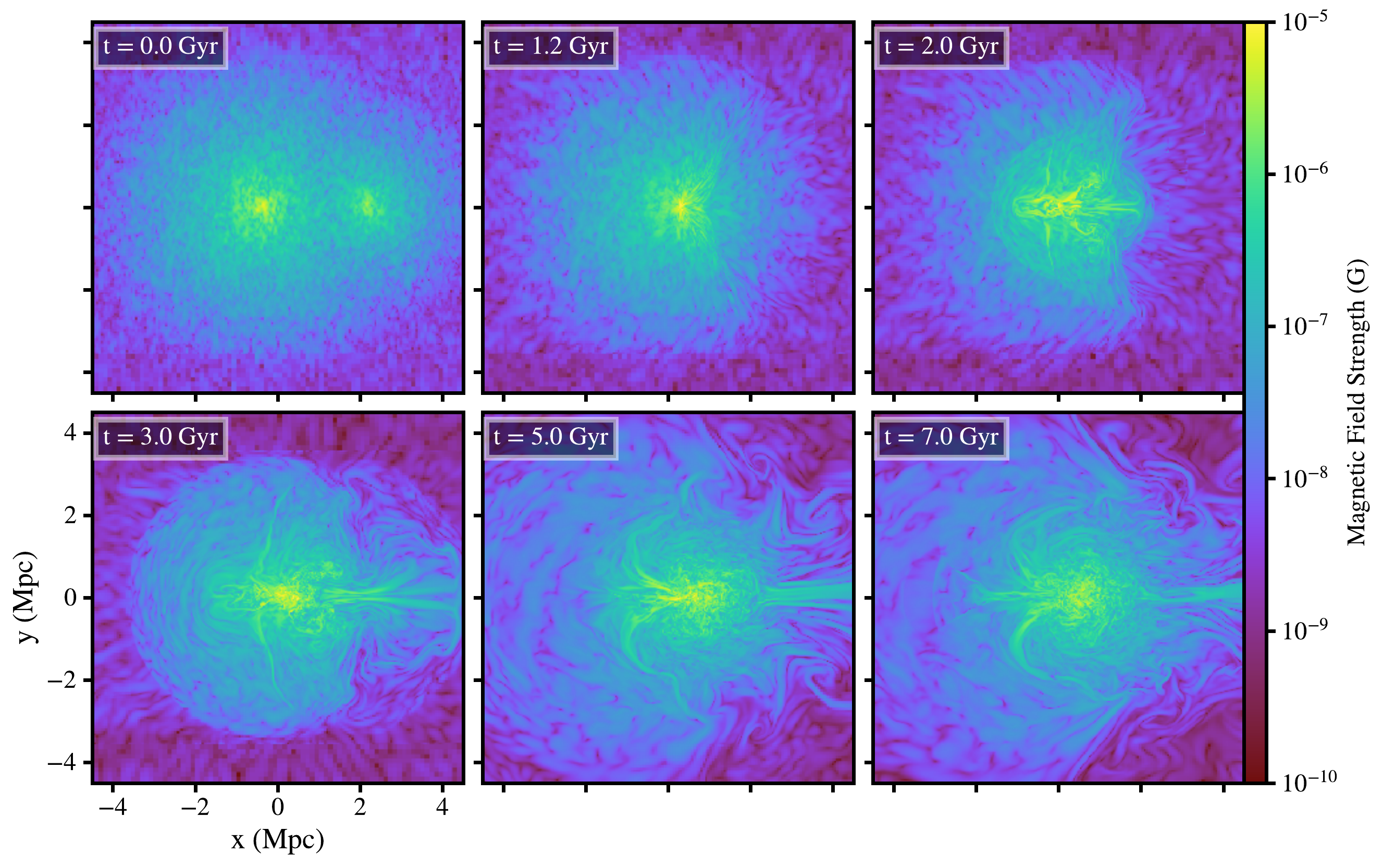}
\end{center}
\caption{Magnetic field strength slices through the collision axis for the MS4 simulation ($M_1/M_2 = 3$, $b = 0$). The epochs shown are: $t$ = 0, 1.2, 2.0, 3.0, 5.0, and 7.0 Gyr.\label{fig:1to3_b0_magfield}}
\end{figure*} 

\begin{figure*}
\begin{center}
\includegraphics[width=0.94\linewidth]{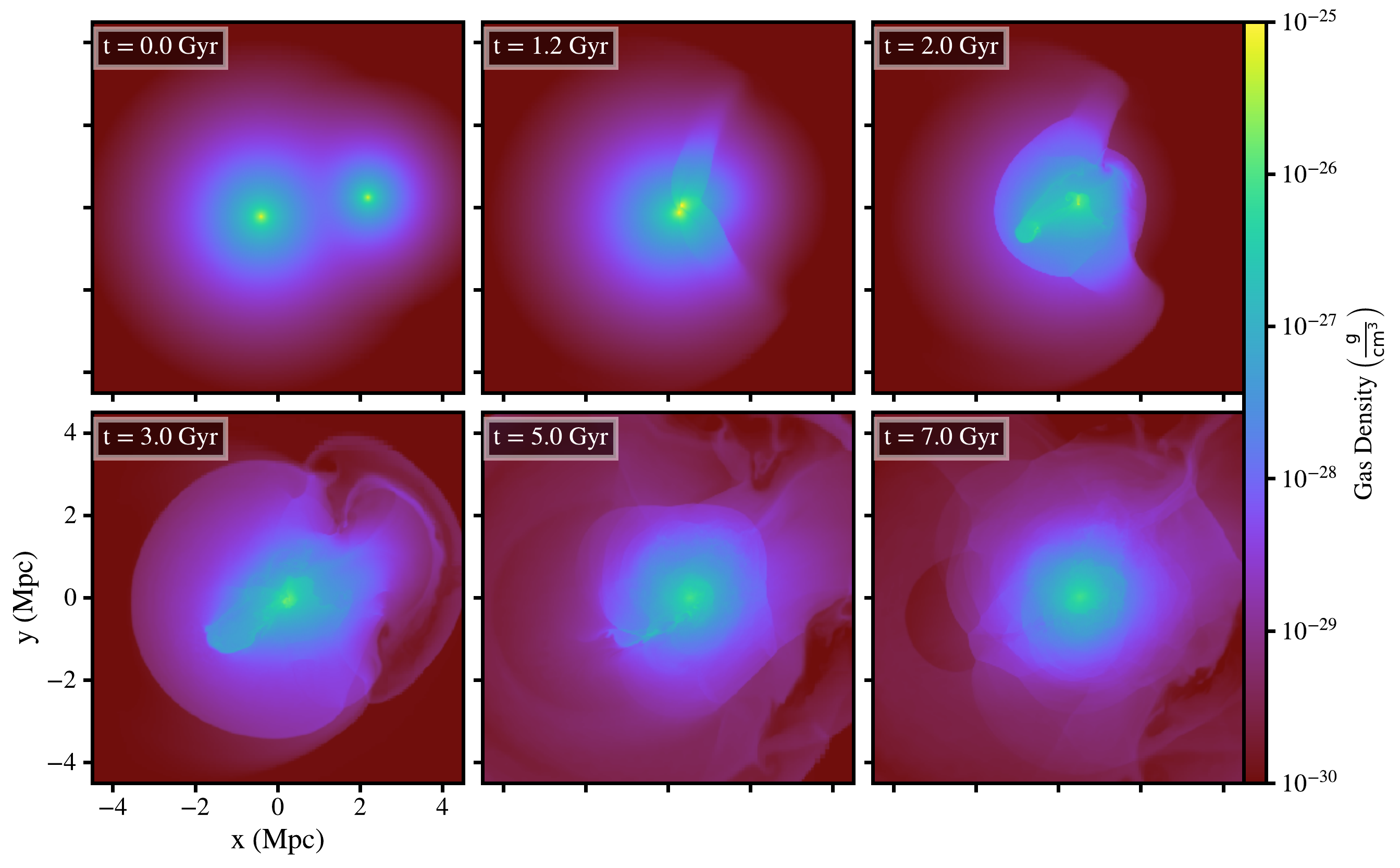}
\end{center}
\caption{Density slices through the collision axis for the MS5 simulation ($M_1/M_2 = 3$, $b = 0.3r_{200}$). The epochs shown are: $t$ = 0, 1.2, 2.0, 3.0, 5.0, and 7.0 Gyr.\label{fig:1to3_b0.5_density}}
\end{figure*} 

\begin{figure*}
\begin{center}
\includegraphics[width=0.94\linewidth]{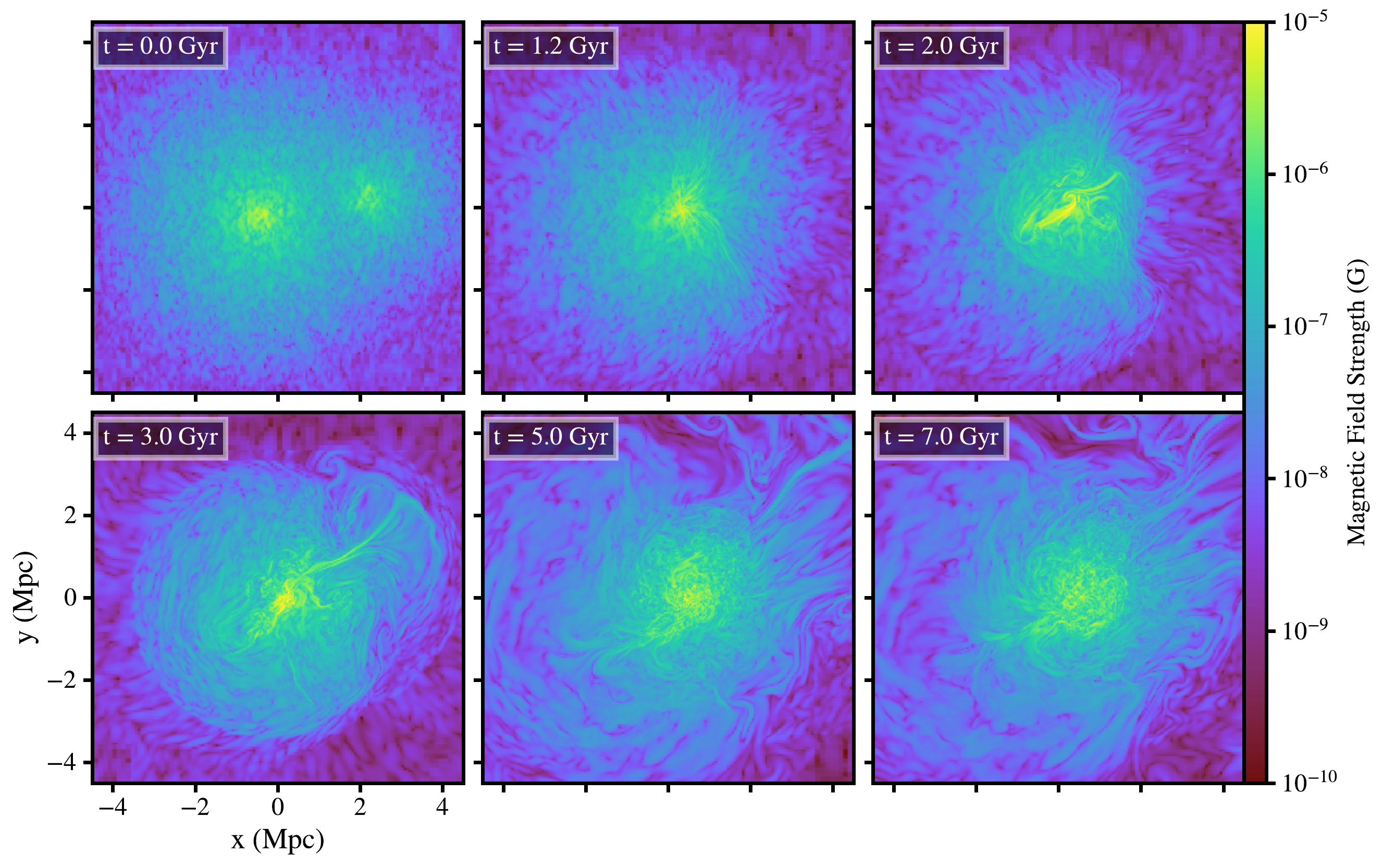}
\end{center}
\caption{Magnetic field strength slices through the collision axis for the MS5 simulation ($M_1/M_2 = 3$, $b = 0.3r_{200}$). The epochs shown are: $t$ = 0, 1.2, 2.0, 3.0, 5.0, and 7.0 Gyr.\label{fig:1to3_b0.5_magfield}}
\end{figure*} 

\begin{figure*}
\begin{center}
\includegraphics[width=0.94\linewidth]{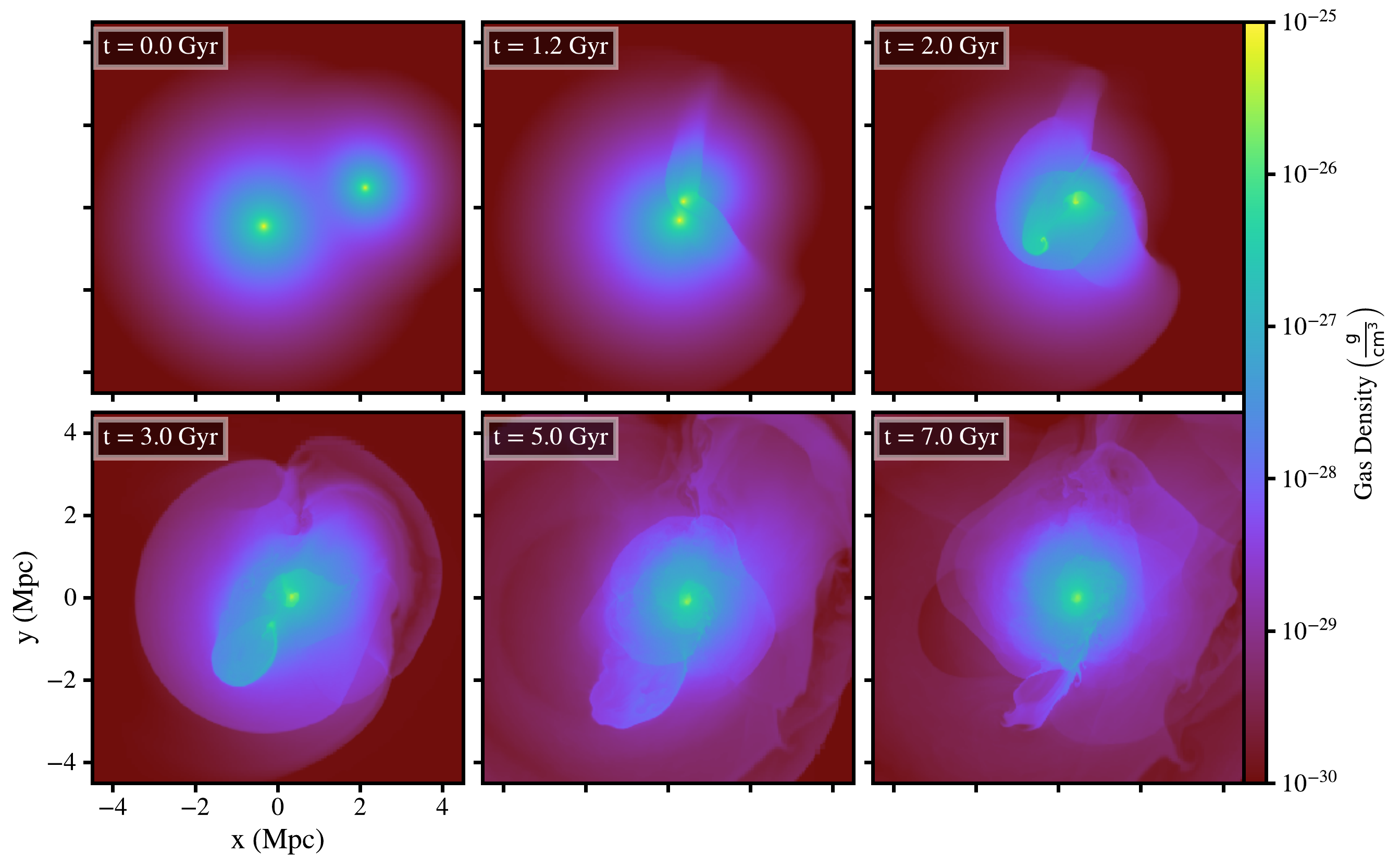}
\end{center}
\caption{Density slices through the collision axis for the MS6 simulation ($M_1/M_2 = 3$, $b = 0.6r_{200}$). The epochs shown are: $t$ = 0, 1.2, 2.0, 3.0, 5.0, and 7.0 Gyr.\label{fig:1to3_b1_density}}
\end{figure*} 

\begin{figure*}
\begin{center}
\includegraphics[width=0.94\linewidth]{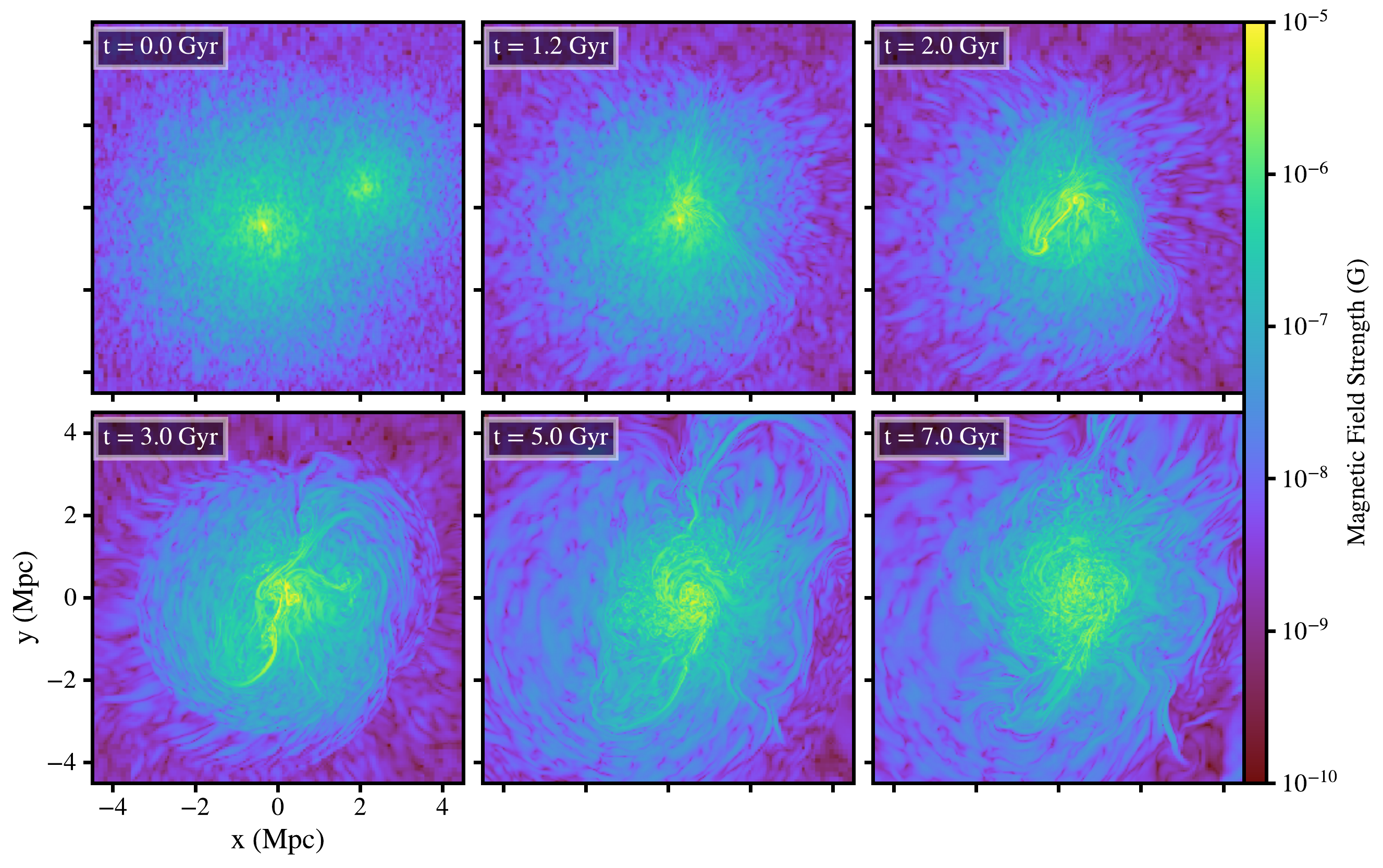}
\end{center}
\caption{Magnetic field strength slices through the collision axis for the MS6 simulation ($M_1/M_2 = 3$, $b = 0.6r_{200}$). The epochs shown are: $t$ = 0, 1.2, 2.0, 3.0, 5.0, and 7.0 Gyr.\label{fig:1to3_b1_magfield}}
\end{figure*} 

\begin{figure*}
\begin{center}
\includegraphics[width=0.94\linewidth]{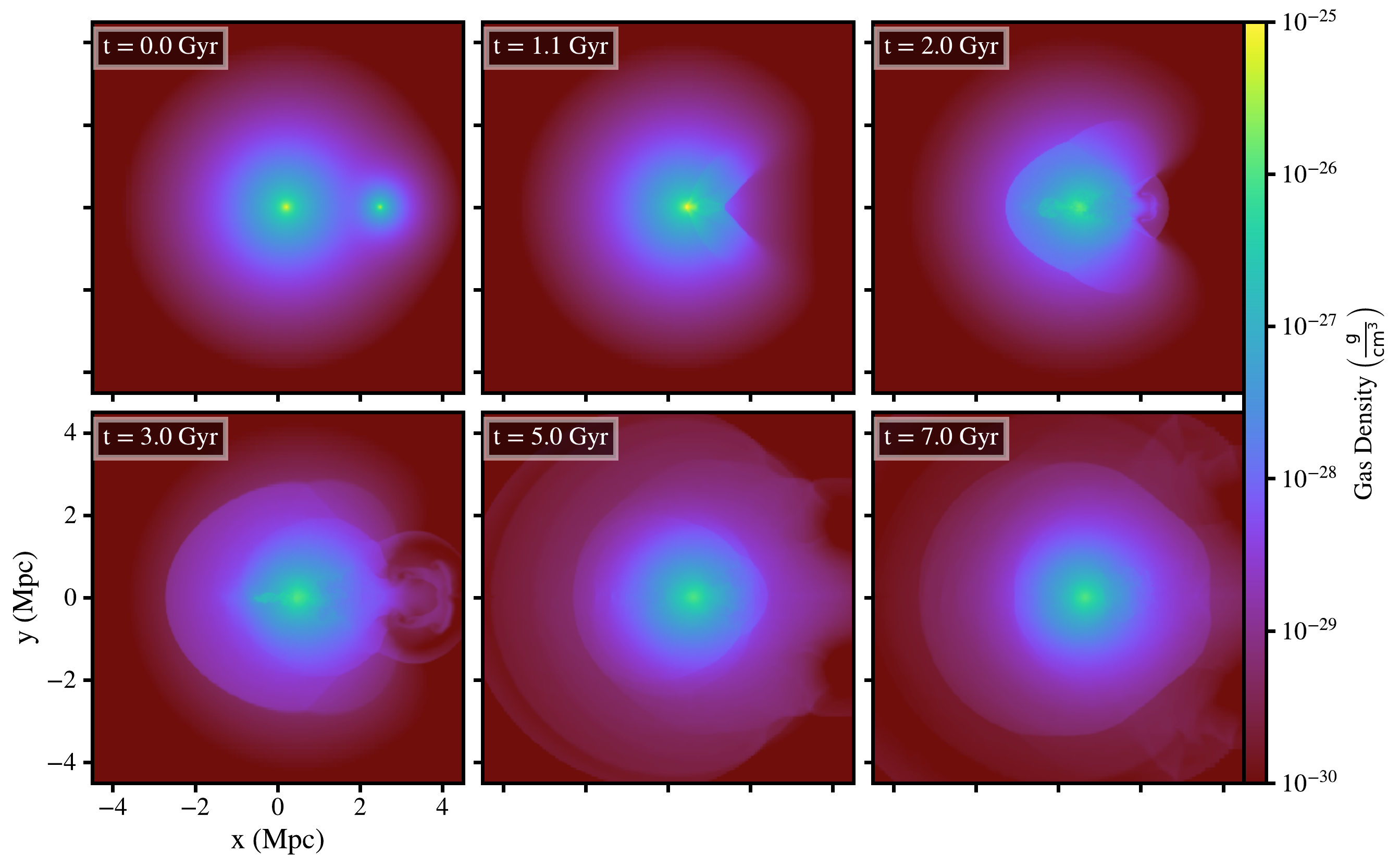}
\end{center}
\caption{Density slices through the collision axis for the MS7 simulation ($M_1/M_2 = 10$, $b = 0$). The epochs shown are: $t$ = 0, 1.1, 2.0, 3.0, 5.0, and 7.0 Gyr.\label{fig:1to10_b0_density}}
\end{figure*} 

\begin{figure*}
\begin{center}
\includegraphics[width=0.94\linewidth]{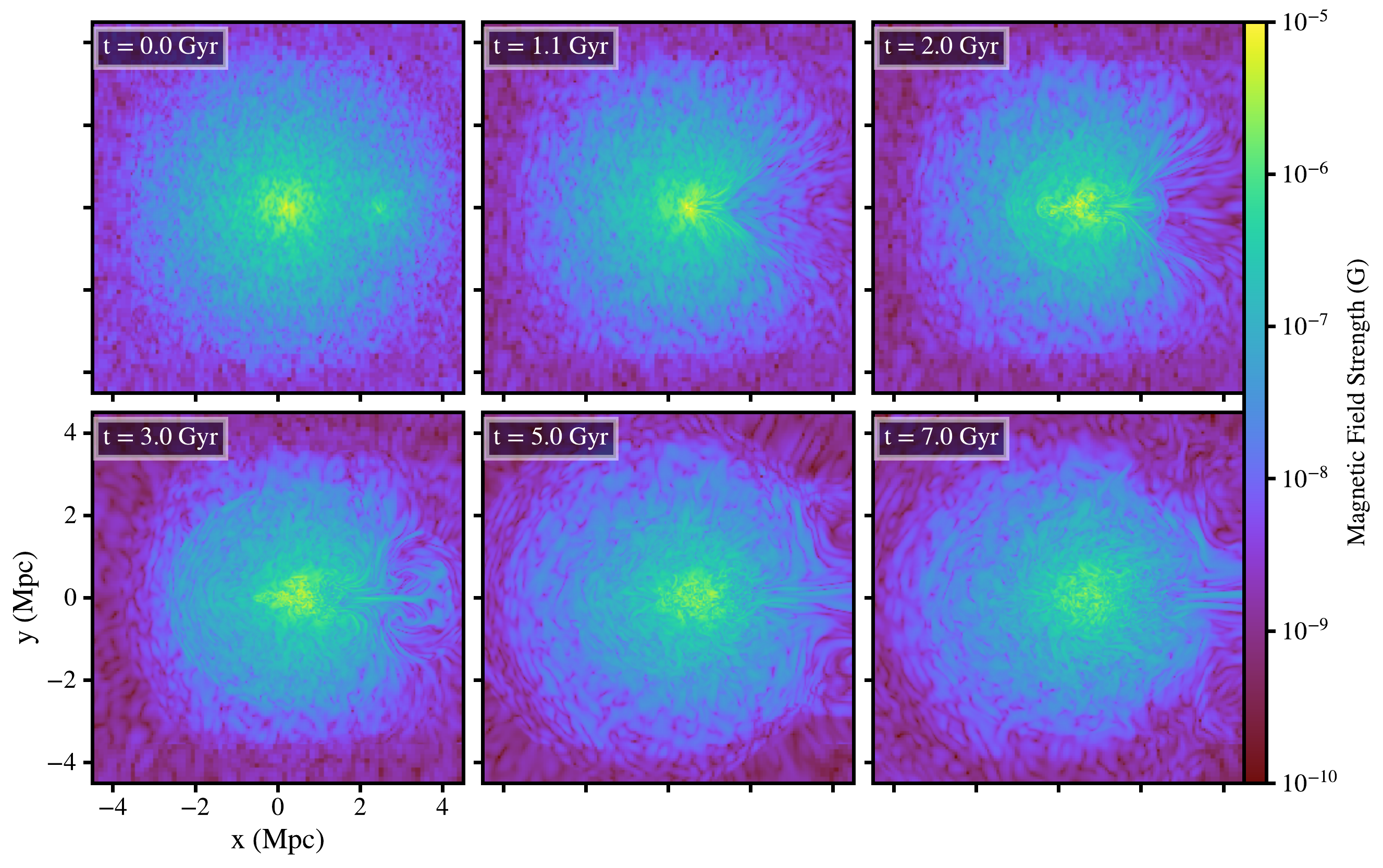}
\end{center}
\caption{Magnetic field strength slices through the collision axis for the MS7 simulation ($M_1/M_2 = 10$, $b = 0$). The epochs shown are: $t$ = 0, 1.1, 2.0, 3.0, 5.0, and 7.0 Gyr.\label{fig:1to10_b0_magfield}}
\end{figure*} 

\begin{figure*}
\begin{center}
\includegraphics[width=0.94\linewidth]{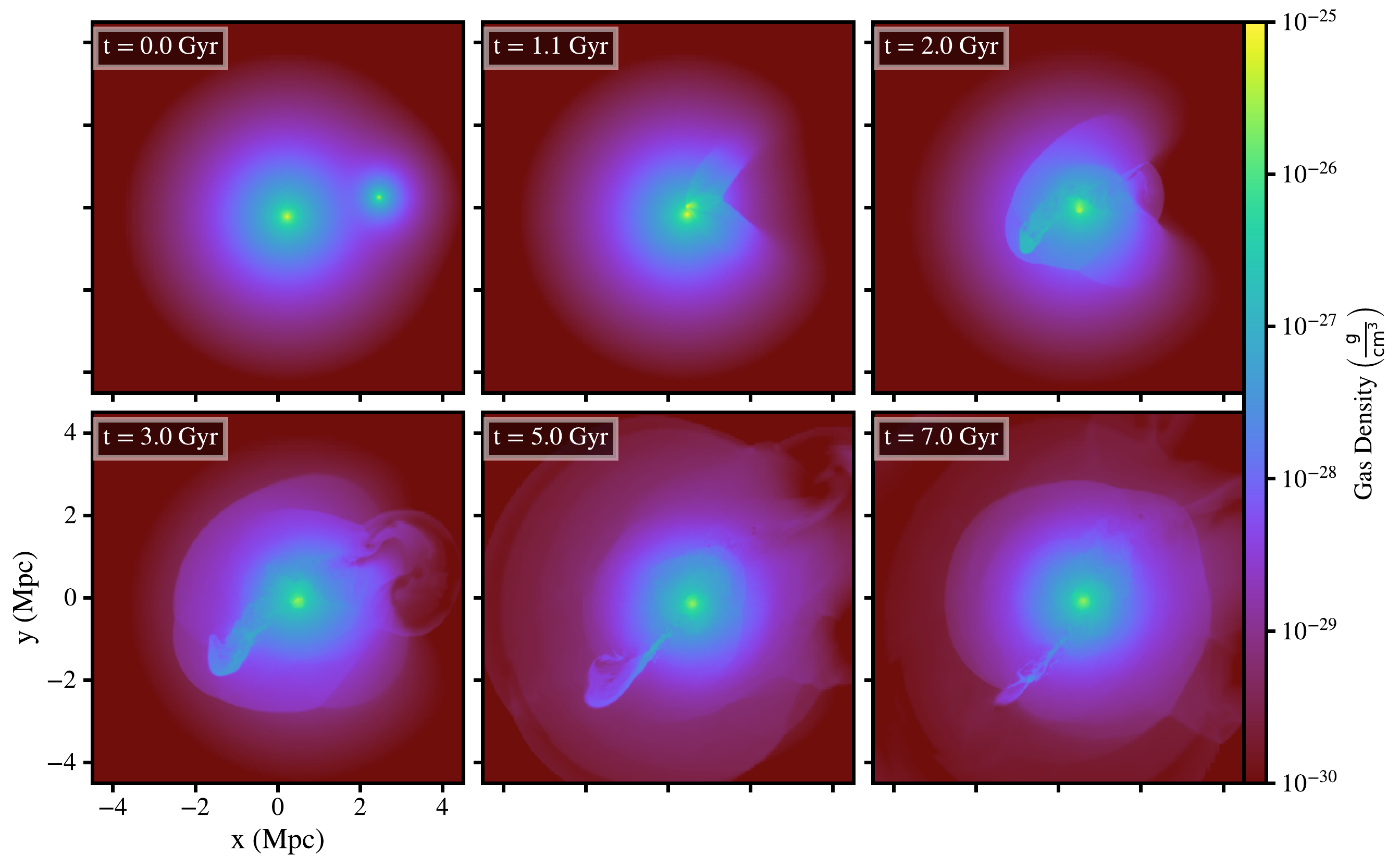}
\end{center}
\caption{Density slices through the collision axis for the MS8 simulation ($M_1/M_2 = 10$, $b = 0.3r_{200}$). The epochs shown are: $t$ = 0, 1.1, 2.0, 3.0, 5.0, and 7.0 Gyr.\label{fig:1to10_b0.5_density}}
\end{figure*} 

\begin{figure*}
\begin{center}
\includegraphics[width=0.94\linewidth]{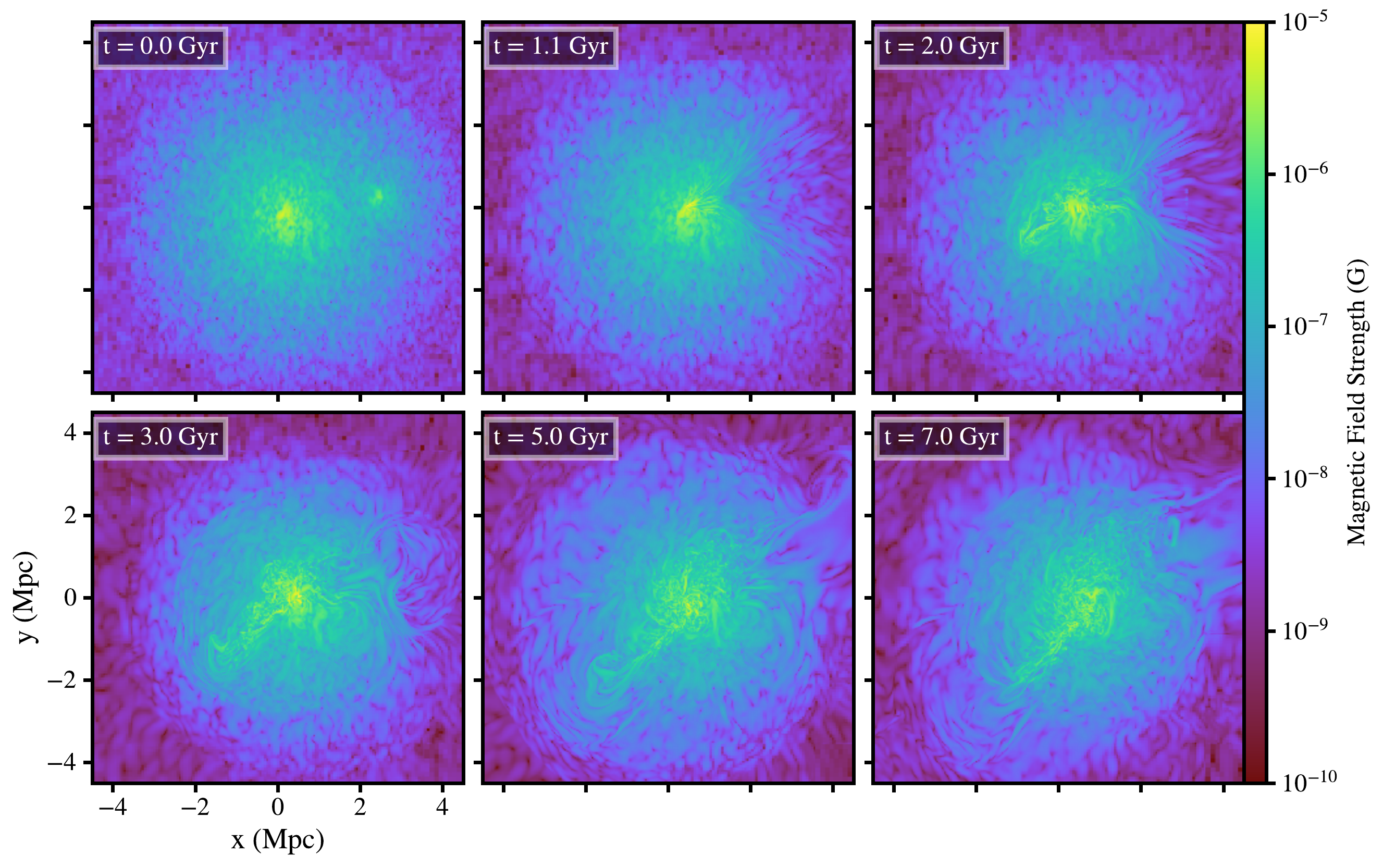}
\end{center}
\caption{Magnetic field strength slices through the collision axis for the MS8 simulation ($M_1/M_2 = 10$, $b = 0.3r_{200}$). The epochs shown are: $t$ = 0, 1.1, 2.0, 3.0, 5.0, and 7.0 Gyr.\label{fig:1to10_b0.5_magfield}}
\end{figure*} 

\begin{figure*}
\begin{center}
\includegraphics[width=0.94\linewidth]{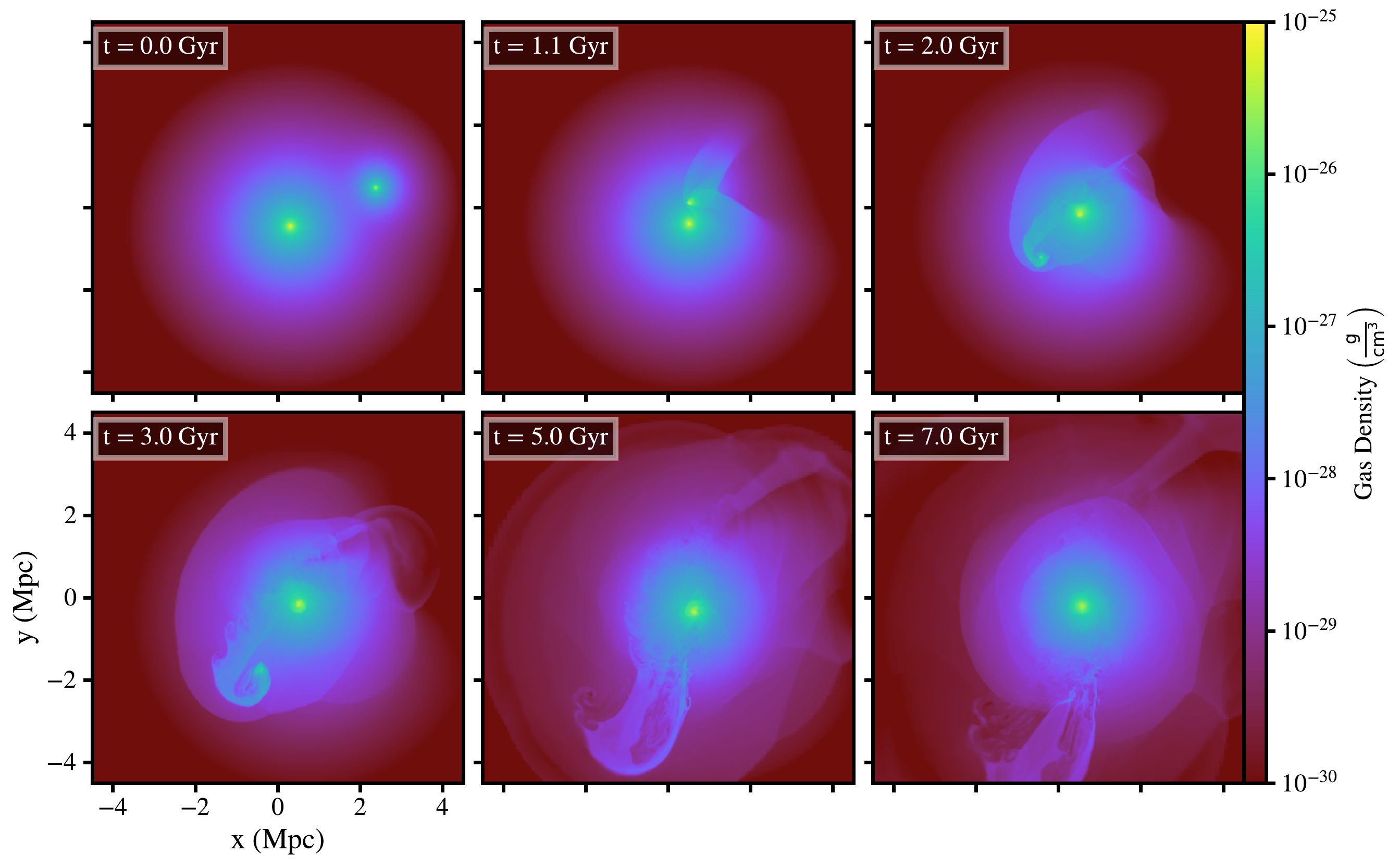}
\end{center}
\caption{Density slices through the collision axis for the MS9 simulation ($M_1/M_2 = 10$, $b = 0.6r_{200}$). The epochs shown are: $t$ = 0, 1.1, 2.0, 3.0, 5.0, and 7.0 Gyr.\label{fig:1to10_b1_density}}
\end{figure*} 

\begin{figure*}
\begin{center}
\includegraphics[width=0.94\linewidth]{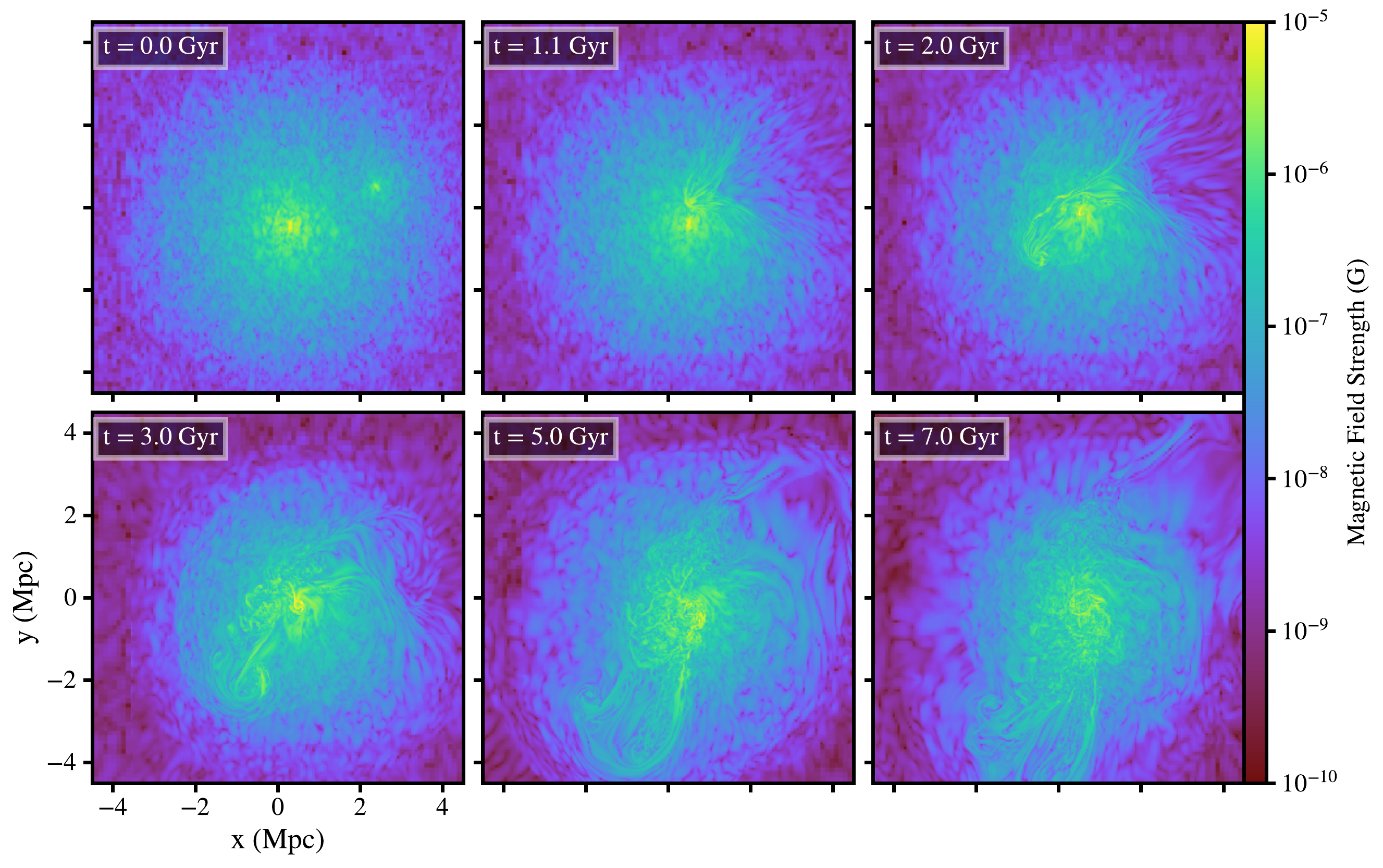}
\end{center}
\caption{Magnetic field strength slices through the collision axis for the MS9 simulation ($M_1/M_2 = 10$, $b = 0.6r_{200}$). The epochs shown are: $t$ = 0, 1.1, 2.0, 3.0, 5.0, and 7.0 Gyr.\label{fig:1to10_b1_magfield}}
\end{figure*} 
  
We first seek to obtain a qualitative picture of how magnetic fields evolve during the different merger stages. Figures \ref{fig:1to1_b0_density}-\ref{fig:1to10_b1_magfield} show slices in density and magnetic field strength perpendicular to the $z$-axis and through the merger plane. In addition to these figures, slices of the temperature and DM density are shown in the Appendix in Figures \ref{fig:1to1_b0_kT}-\ref{fig:1to10_b1_all_cic}. The evolution of the gas thermodynamical properties within each cluster merger for our simulations is essentially identical to that in the simulations in Z11, and have also been described in previous binary merger simulation investigations \citep[e.g.][]{ric01,poo06}. Here, we focus on the evolution of the magnetic field structures during the merger. Many of the magnetic field structures we note below were noticed in similar simulations in \citet{roe99} and \citet{tak08}, though we have many more combinations of mass ratio and impact parameter than those earlier works. 
    
For each simulation MS1-MS9, we make slices at 6 different points in time. Time $t = 0$~Gyr is the initial state, where the virial radii of both clusters are just touching. The next epoch is chosen to approximately mark the first core passage for each simulation. In MS1-MS3, this is at about $t\approx1.4$ Gyr; in MS4-MS6, $t\approx1.2$ Gyr; in MS7-MS9, $t\approx1.1$ Gyr. After that, the plotted epochs are $t$ = 2.0, 3.0, 5.0, and 7.0~Gyr. These epochs are simply chosen to show relevant epochs from the first two core passages until late in the simulation. 

The equal-mass mergers have a qualitatively different evolution from the unequal-mass cases, so we will discuss each of these separately.

\subsubsection{Equal-mass Mergers}
  
In the $M_1/M_2 = 1, b = 0$ merger (simulation MS1, Figure \ref{fig:1to1_b0_density}), the first core passage occurs at about $t = 1.4$ Gyr. Building up to the core passage, the gas in the midplane of the merger becomes shocked and compressed and forms a flat, ``pancake''-like structure perpendicular to the line of centers between the cluster cores. The compressed gas amplifies the magnetic field in these regions. The magnetic field strength is also increased behind the shock fronts as they propagate outward. The DM cores, unimpeded by ram pressure, pass through and then subsequently oscillate about each other. These rapid and violent changes in the gravitational potential drag the surrounding gas back and forth, driving smaller shocks and turbulence. At first, these motions create long, thin, and laminar strong ``filament'' structures in the field along the merger axis with strengths up to $\sim$10~$\mu$G (Figure \ref{fig:1to1_b0_magfield}), but as the gas is violently stirred by the DM cores, a turbulent magnetic field is generated in the center. Turbulent fields are also generated from the gravitational infall of the pancake structure into the center. 
  
\begin{figure*}[tp!]
\begin{center}
\includegraphics[width=0.95\linewidth]{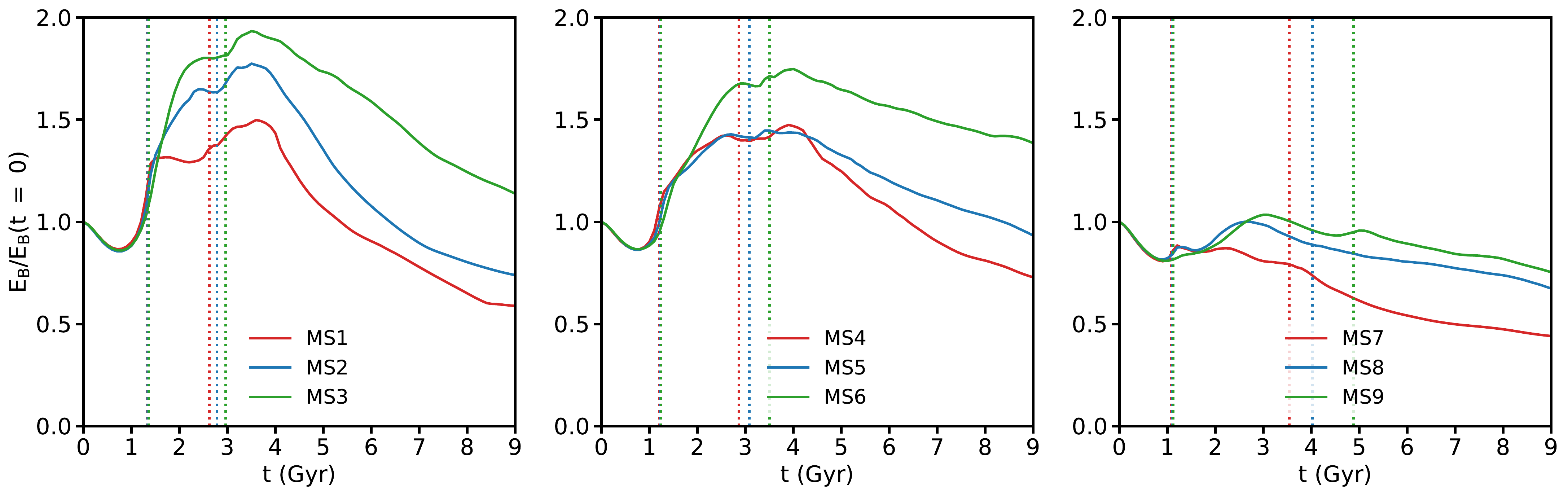}
\caption{Magnetic energy vs. time for all the simulations within the central volume of $V = (8~{\rm{Mpc}})^3$, scaled by the value at $t = 0$~Gyr. Left panels: $M_1/M_2 = 1$ simulations. Red corresponds to $b = 0$, blue to $b = 0.3r_{200}$, and green to $b = 0.6r_{200}$. Center: $M_1/M_2 = 3$ simulations. Right: $M_1/M_2 = 10$ simulations. The vertical dashed lines correspond to the first and second perigee passages for each merger, and have the same colors as the simulations listed in the legend.\label{fig:emag_evol}}
\end{center}
\end{figure*} 
  
In the $M_1/M_2 = 1$ mergers with nonzero impact parameter (simulations MS2 and MS3, Figures \ref{fig:1to1_b0.5_density}-\ref{fig:1to1_b1_magfield}), the gas and DM cores ``sideswipe,'' ram-pressure stripping gas through the two cores. Two cold fronts $\sim$1-2~Mpc in length emerge from the center at $t \sim 3$~Gyr, expanding mostly radially outward from their respective cores. The magnetic field is stretched and amplified along these fronts by their associated velocity shears up to $\sim$10~$\mu$G, which produces the same ``filament'' structures in the field as seen in the $M_1/M_2 = 1, b = 0$ merger simulation. These structures are short-lived, however, as the DM cores undergo their second and third core passages, driving yet more turbulence. Similar to the $M_1/M_2 = 1, b = 0$ case, long, straight field structures can persist for longer at larger radii $r \simgt 1.5$~Mpc. 

In all of the $M_1/M_2 = 1$ simulations, at large radii ($r \simgt 1.5$~Mpc), long, mostly straight magnetic field lines are also stretched and amplified between gas which is moving radially outward, on either side of the cores, and gas which is falling back into them. These structures, Mpc in length, persist until the end of each simulation at $t = 10$~Gyr, but within $r \sim 1.5$~Mpc random turbulent motions have produced a mostly turbulent magnetic field.

\subsubsection{Unequal-mass Mergers}

At first core passage in the unequal-mass, $b = 0$ mergers (simulations MS4 and MS7, Figures \ref{fig:1to3_b0_density}, \ref{fig:1to3_b0_magfield}, \ref{fig:1to10_b0_density}, and \ref{fig:1to10_b0_magfield}), the core of the secondary punches through that of the primary, completely disrupting it. At the same time, gas is ram-pressure stripped from the secondary and mixed in with the primary's gas via Kelvin-Helmholtz instabilities. Roughly 2~Gyr after the first core passage (t = 3.0~Gyr), the remaining core gas from the secondary (and its DM core as well) begin to fall back into the secondary, forming a cold inflow of gas that is shock-heated as it enters the core region. As the secondary cluster passes through the primary and returns back, it stretches the magnetic field behind it via shear amplification into similar filament features. As both gas from the primary and the stripped gas from the secondary start to fall back onto the oscillating cores, the magnetic field lines wrap inwards. These magnetic field structures are initially fairly laminar, but quickly become tangled within the core region due to Kelvin-Helmholtz instabilities and turbulence.
  
In the non-zero impact parameter, unequal mass cases (simulations MS5, MS6, MS8, and MS9), the primary cluster core produces sloshing cold fronts, and the secondary cluster develops a dense, cold, and long ($\sim$~a few Mpc) plume of gas which trails behind it and is stripped as it leaves the primary's core region and later returns, also producing a cold front (Figures \ref{fig:1to3_b0.5_density}, \ref{fig:1to3_b1_density}, \ref{fig:1to10_b0.5_density}, and \ref{fig:1to10_b1_density}). Along these cold fronts and plume, the magnetic field is similarly stretched and amplified as in the equal-mass, non-zero impact parameter cases (Figures \ref{fig:1to3_b0.5_magfield}, \ref{fig:1to3_b1_magfield}, \ref{fig:1to10_b0.5_magfield}, and \ref{fig:1to10_b1_magfield}). Because the second core passage occurs much later in these simulations, these magnetic field structures can persist for a few Gyr longer than in the other simulations.

\subsection{The Evolution of the Magnetic Energy Over Time}\label{sec:energyovertime}
  
Figure \ref{fig:emag_evol} shows the evolution of the magnetic energy within the central $V = (8~{\rm{Mpc}})^3$ in the simulations, scaled by the value of the magnetic energy at $t = 0$~Gyr. This volume is large enough to contain the two clusters out to their respective $r_{200}$ initially and their subsequent evolution. The first and second core passages are marked in each panel with vertical dashed lines with the same colors as the lines in the legend.

The stages in the evolution of the magnetic energy can be explained by reference to the events detailed in Section \ref{subsec:slices}. After a short period of decline as the initial field relaxes, it increases rapidly in all simulations at the first core passage due to compression of the gas. A slower increase after the core passage occurs due to stretching of the field lines due to shear motions, which are more significant for larger impact parameters. After this, the magnetic energy increase flattens out, as the cluster cores move away from each other and the gas re-expands. Another increase occurs at the second core passage, with more gas compression and stretching of field lines. Throughout the period between the first and second core passages, additional energy is transferred to the magnetic field through the stretching of field lines from turbulence. The increase in the magnetic energy is less significant in simulations with smaller subclusters, but for a given mass ratio it is more significant for larger impact parameters, as the shearing motions generated after the first core passage strongly amplify the magnetic field.

At the end of every simulation, the magnetic energy gradually decreases, as the magnitude of the compression and stretching diminishes. We have verified that a negligible amount of magnetic energy is being advected through the boundaries of our $V = (8~{\rm{Mpc}})^3$ volume, thus the main reasons for the decrease in the magnetic energy after the second core passage are the expansion of the core due to the increase of its entropy (see Section \ref{subsec:radialprofiles}) and the relaxation of the magnetic tension in the turbulent field. This late-time decrease in magnetic field strength appears to be in conflict with the results of previous simulations of cluster mergers (idealized or cosmological) with magnetic fields, where the magnetic field increases continuously and gradually without decreasing \citep[e.g.][]{dolag2002,xu2009,marinacci2015,vazza2018,domfern19}. Needless to say, our binary merger simulations are very different in the respect that only one merger between two clusters is happening, whereas in cosmological simulations there is constant merging and accretion of material. Sustaining such an increase in the magnetic energy requires a setting where the drivers of turbulent and compressive motions are being constantly replenished \citep{sub06}. Our simulations also lack radiative cooling, which would compress the core gas and amplify the field in these regions after the effects of the merger subside. In \citet{roe99}, another work involving idealized mergers such as ours, the magnetic field does increase for 5~Gyr after the first core passage, but the low spatial resolution in that work may have prevented the gas mixing that drives the expansion of the core gas and hence decreases the magnetic field in the core (see their Figure 5 and Section \ref{subsec:radialprofiles} of this work). 
  
\subsection{Radial Profiles of the Final State} \label{subsec:radialprofiles}

By the end of each simulation ($t$ = 10~Gyr), the clusters have fully merged and are nearly relaxed. It is instructive to  examine radial profiles of the gas properties. For our purposes, the most relevant quantities to examine are those related to the velocity and magnetic fields, as well as the gas entropy. The radial profiles are taken in radial bins centered on the cluster potential minimum.
    
\subsubsection{Radial Profiles of Velocity Fields}\label{sec:velocity_profiles}

The local velocity dispersion is a measurement of the turbulent kinetic energy:
\begin{eqnarray}
\sigma_v^2 &=& \langle v^2\rangle-\langle v\rangle^2
\end{eqnarray}
We want to compare the relative contributions to the energy from the internal (IE), kinetic (KE), and magnetic (ME) energies, so we also compute profiles of the sound and Alfv\'en speeds:
\begin{eqnarray}
c_s^2 &=& \frac{\gamma{P}}{\rho} \\
v_A^2 &=& \frac{\langle B^2\rangle}{4\pi\rho},
\end{eqnarray}
where $\gamma=5/3$. These are the characteristic speeds for the internal and magnetic energies, respectively. Taking ratios of these squared velocities essentially yields the ratios between these different forms of energy:
\begin{eqnarray} \label{eqn:energy_ratios}
\frac{\sigma_v^2}{c_s^2} &=& {\cal M_{\rm turb}}^2 = \frac{\sigma_v^2}{\gamma{P}/\rho} = \frac{\rho\sigma_v^2/2}{\gamma(\gamma-1)\rho{\epsilon}} \sim \frac{\rm KE}{\rm IE} \\
\frac{\sigma_v^2}{v_A^2} &=& {\cal M_{\rm A,turb}}^2 = \frac{\sigma_v^2}{{\langle B^2\rangle}/{4\pi\rho}}=\frac{\rho\sigma_v^2/2}{\langle B^2\rangle/8\pi} = \frac{\rm KE}{\rm ME}
\end{eqnarray} 
where ${\cal M}_{\rm turb}$ ${\cal M}_{\rm A,turb}$ are the Mach and Alfv\'enic Mach numbers of the turbulent gas motions, respectively. These ratios quantify the physical relevance of one type of energy to another. 
  
\begin{figure*} [tp!]
\begin{center}
\includegraphics[width=\linewidth]{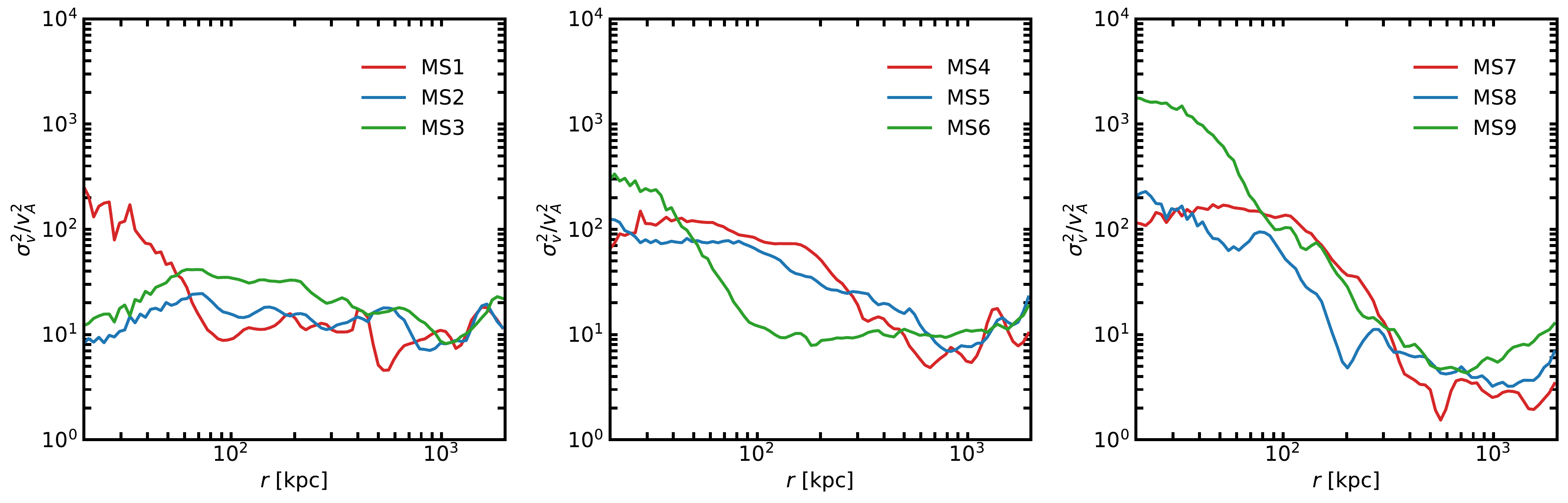}
\end{center}
\caption{Radial profiles of $\sigma_v^2/v_A^2$. Left: Profiles for the $M_1/M_2 = 1$ simulations. Red corresponds to $b = 0$, blue to $b = 0.3r_{200}$, and green to $b = 0.6r_{200}$. Center: Profiles for $M_1/M_2 = 3$. Right: Profiles for $M_1/M_2 = 10$.\label{fig:sigma_vA}}
\end{figure*}

\begin{figure*} [tp!]
\begin{center}
\includegraphics[width=\linewidth]{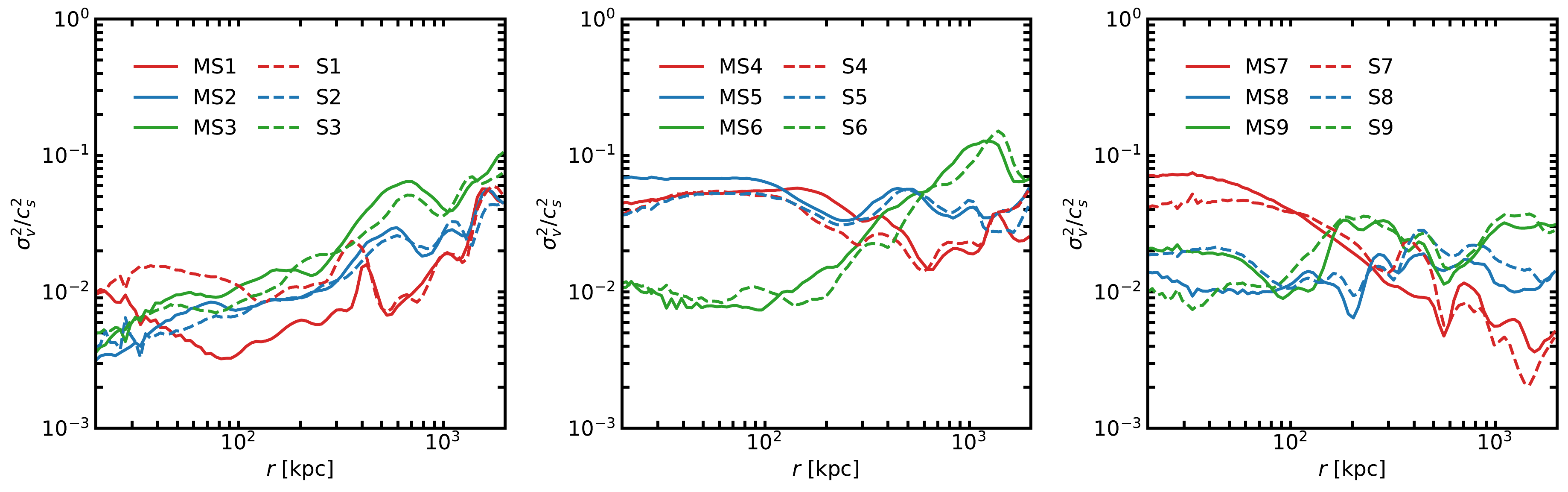}
\end{center}
\caption{Radial profiles of $\sigma_v^2/c_s^2$. Left: Profiles for the $M_1/M_2 = 1$ simulations. Solid lines are magnetized simulations, dashed are unmagnetized. Red corresponds to $b = 0$, blue to $b = 0.3r_{200}$, and green to $b = 0.6r_{200}$. Center: Profiles for $M_1/M_2 = 3$. Right: Profiles for $M_1/M_2 = 10$.\label{fig:sigma_cs}}
\end{figure*}

Figures \ref{fig:sigma_vA} and \ref{fig:sigma_cs} show profiles of these ratios at the final state of all of our simulations. First, Figure \ref{fig:sigma_vA} shows profiles of $\sigma_v^2/v_A^2$. Within the core region of $r \sim 100-300$~kpc, this ratio is generally on the order of $\sigma_v^2/v_A^2 \sim 10-100$. In the equal mass mergers, increasing impact parameter has only a mild effect on this ratio, since each case is exceptionally turbulent. In the equal-mass, zero-impact parameter case, given the symmetry of the simulation, the gas components of two clusters do not significantly penetrate one anothers' atmospheres but instead ``stick together''. Hence, there is not nearly as much shear amplification in this case, and the magnetic field is weaker in the center than in the equal-mass, off-center collisions. In the high mass ratio mergers, there is a mild trend towards higher ratios of $\sigma_v^2/v_A^2$ with increasing impact parameter, since in those cases, turbulence has gone more into heating the core and expanding it (thus lowering the magnetic field strength) than into amplifying and stretching the magnetic field. For example, in MS9, the magnetic field has only been mildly amplified by turbulence since the subcluster is small and passes by at a large distance. Though there is still turbulence in the core, it transfers little energy to the magnetic field. This result is consistent with that of Section \ref{sec:energyovertime}. 
  
Beyond the core, this ratio is closer to $\sim$10 or even lower, with the magnetic energy more comparable to the kinetic energy, but still lower. It should be noted again that our idealized simulations began with no turbulent velocities but with a turbulent magnetic field, highlighting the efficiency with which the bulk motion of the merger is converted into turbulent motion and the relative inefficiency with which that energy is transferred to the magnetic field \citep[see also][]{min15,vazza2018}.

Figure \ref{fig:sigma_cs} shows profiles of $\sigma_v^2/c_s^2$ for the magnetized and unmagnetized simulations. In all cases, there are only modest differences in the turbulent kinetic energy profiles between the magnetized and unmagnetized simulations, again consistent with the results of Section \ref{sec:energyovertime}.

\subsubsection{Radial Profiles of Entropy}\label{sec:entropy_profiles}

Z11 showed that turbulent mixing from cluster mergers can increase the entropy of the core region substantially for a range of mass ratios and impact parameters. Magnetic fields are capable of preventing this mixing if they are strong enough (see Figure 24 of ZML11 for an illustration of this effect in a sloshing cluster core). If this effect is strong, we expect the core entropies in the magnetic simulations to be significantly lower than their unmagnetized counterparts.

Figure \ref{fig:entropy} shows the entropy profiles for the magnetized and unmagnetized simulations. In each case, the magnetized simulations have slightly lower core entropies than the corresponding unmagnetized simulations in most cases, but this effect is very modest. Given that our previous results have shown that the magnetic field has little effect on the turbulent velocities in our merger simulations compared to their unmagnetized versions from Z11, this should come as no surprise. As was seen in Z11, higher levels of core entropy tend to be correlated with higher turbulence in the core region, as seen in Figure \ref{fig:sigma_cs}.
  
\subsubsection{Radial Profiles of Magnetic Field Strength}\label{sec:bfield_profiles}
  
\begin{figure*} [tp!]
\begin{center}
\includegraphics[width=\linewidth]{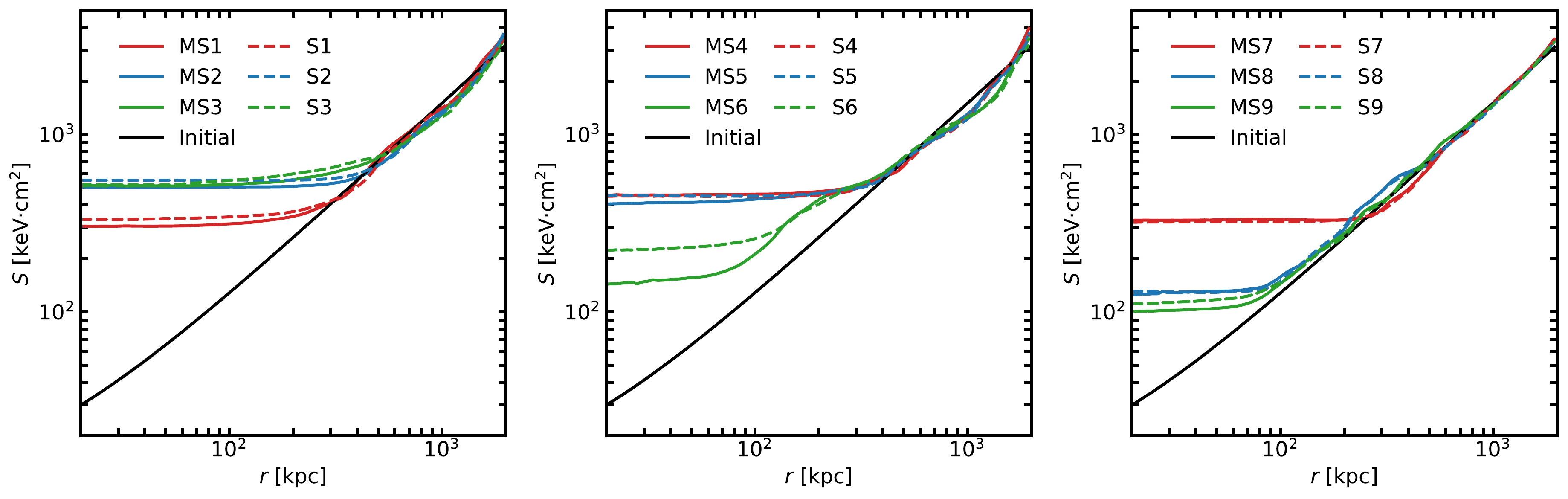}
\end{center}
\caption{Radial profiles of entropy $S$. Left: Profiles for the $M_1/M_2 = 1$ simulations. Solid lines are magnetized simulations, dashed are unmagnetized. Red corresponds to $b = 0$, blue to $b = 0.3r_{200}$, and green to $b = 0.6r_{200}$. Center: Profiles for $M_1/M_2 = 3$. Right: Profiles for $M_1/M_2 = 10$.\label{fig:entropy}}
\end{figure*}

\begin{figure*} [t!]
\begin{center}
\includegraphics[width=\linewidth]{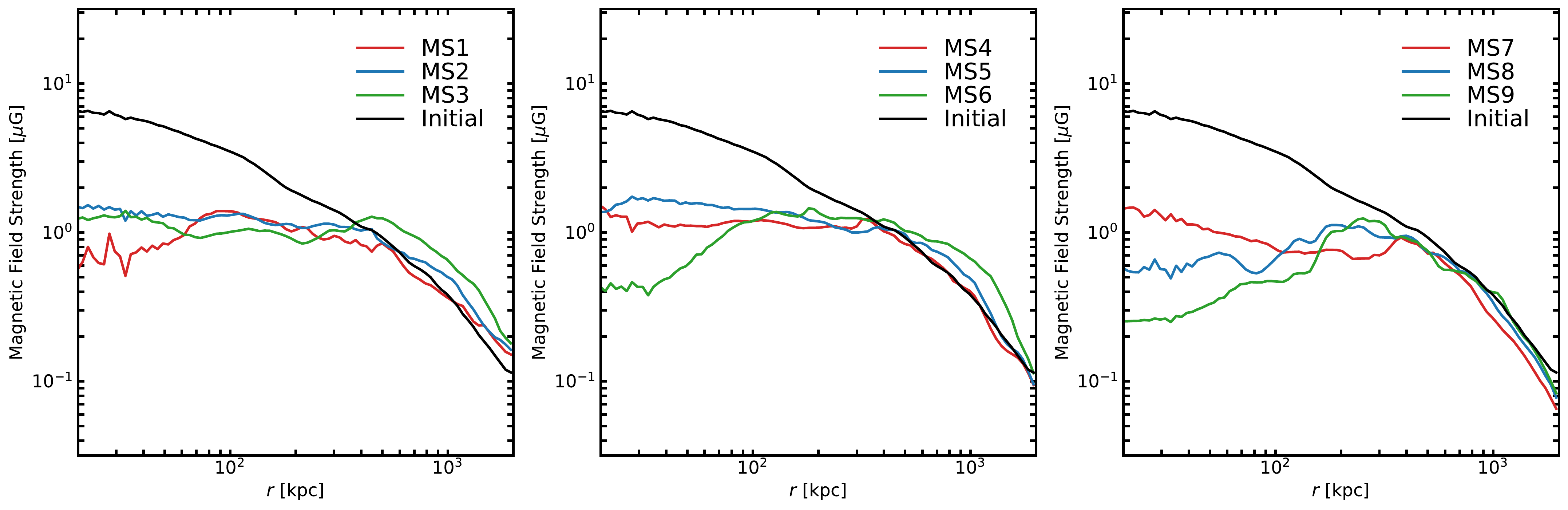}
\end{center}
\caption{Radial profiles of the magnetic field strength at the initial and final states for all simulations. Left: Profiles for the $M_1/M_2 = 1$ simulations. Red corresponds to $b = 0$, blue to $b = 0.3r_{200}$, and green to $b = 0.6r_{200}$. Center: Profiles for $M_1/M_2 = 3$. Right: Profiles for $M_1/M_2 = 10$.\label{fig:magfield_final}}
\end{figure*}

Figure \ref{fig:magfield_final}, shows the magnetic field strength at the initial and final states of each simulation. Within the inner region where the entropy is constant with radius (Figure \ref{fig:entropy}), the magnetic field strength is an order of magnitude or more weaker than that in the initial condition. In these regions, turbulent mixing has lowered the gas density considerably, flattening the core (see Figure 15 from Z11), and due to the effect of flux freezing this causes the magnetic energy to decrease. In the large mass ratio, large impact parameter simulations, the magnetic field strength decreases steeply towards the center of the cluster, because there is not enough turbulence to mix the gas and produce large isentropic cores, and not enough of this energy goes into amplifying the magnetic field in these regions. Outside of the core regions, the magnetic field strength has instead increased, with larger values for smaller mass ratios/larger secondary clusters and larger impact parameters. 

\subsubsection{Radial Profiles of Velocity and Magnetic Field Anisotropy}\label{sec:beta_profiles}

Lastly, we examine the anisotropy of the velocity and magnetic fields at the final state. To do this, we employ a common paramerization for the anisotropy in spherical coordinates:
\begin{eqnarray}
\beta_v &=& 1-\frac{\sigma_{vt}^2}{2\sigma_{vr}^2} \\
\beta_B &=& 1-\frac{\sigma_{Bt}^2}{2\sigma_{Br}^2}
\end{eqnarray}
where $\sigma_{vr}^2$ and $\sigma_{Br}^2$ are the variances of the radial components of the velocity and magnetic fields, respectively, and $\sigma_{vt}^2 = \sigma_{v\theta}^2 + \sigma_{v\phi}^2$ and $\sigma_{Bt}^ 2= \sigma_{B\theta}^2 + \sigma_{B\phi}^2$ are the same for the tangential components. With this parameterization, perfect isotropy of a vector field is $\beta = 0$, preferentially radial vectors have $1 \geq \beta > 0$, and preferentially tangential vectors have $\beta < 0$. 

\begin{figure*} [t!]
\begin{center}
\includegraphics[width=\linewidth]{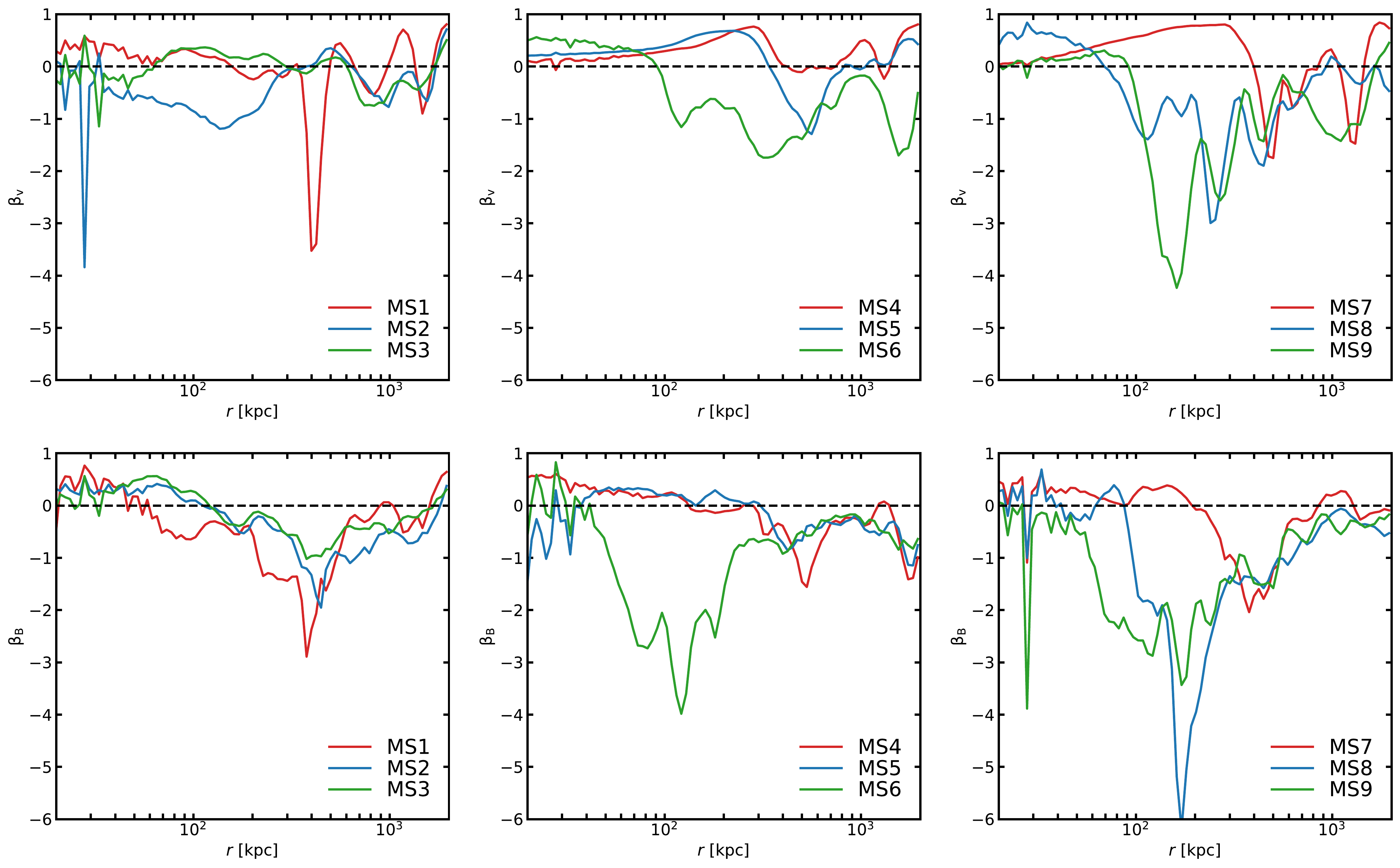}
\end{center}
\caption{Radial profiles of the velocity (top panels) and magnetic field (bottom panels) anisotropies at the final state for all simulations. Left: Profiles for the $M_1/M_2 = 1$ simulations. Red corresponds to $b = 0$, blue to $b = 0.3r_{200}$, and green to $b = 0.6r_{200}$. Center: Profiles for $M_1/M_2 = 3$. Right: Profiles for $M_1/M_2 = 10$. The horizontal black line at $\beta = 0$ indicates perfect isotropy of the given vector field.\label{fig:beta_final}}
\end{figure*}

Figure \ref{fig:beta_final} shows the velocity and magnetic field anisotropy radial profiles for the final state. In the core regions, both the velocity and magnetic field anisotropies are very nearly isotropic or mildly radially anisotropic. In these regions, the cores are isentropic and not stratified, and thus bouyancy effects do not prevent the radial gas motions from becoming as large as the tangential ones. The $b = 0$ simulations (MS1, MS4, MS7) have more radial anisotropy ($\beta > 0$) within and outside of this core region due to the fact that merger proceeds along an entirely radial trajectory. The magnetic field strength is less radially anisotropic than the velocity, however.

Conversely, there is a considerable amount of tangential anisotropy ($\beta < 0$) in both fields in the simulations with nonzero $b$. This is most noticeable in the large mass ratio simulations MS6, MS8, and MS9. There are two reasons for this effect. These simulations have high angular momentum in the initial merger configuration, and the resulting tangential gas motions will stretch the magnetic field lines in a more tangential direction. 

In general, then, cluster mergers produce turbulent magnetic fields which are either mostly isotropic or have a tangential bias. Since conductive heat fluxes are parallel to magnetic field lines in the ICM due to the previously mentioned fact that $\lambda_{\rm mfp} \gg \rho_{\rm L}$, the isotropic and tangential magnetic fields caused by merger-driven turbulence will more efficiently prevent thermal conduction from hotter regions in the outer cluster to the core region in a cool-core cluster than radial fields would \citep{par10,rus10}. As noted in previous works, if the thermal conductivity of the ICM parallel to magnetic field lines is efficient, cool-core clusters with positive temperature gradients are susceptible to the heat-flux buoyancy instability, which can rearrange magnetic fields tangentially and thereby suppress radial thermal conduction \citep{qua08, par08,par09,bog09}. On the other hand, the gas in the outer regions of the cluster are susceptible to the magnetothermal instability, which rearranges magnetic fields radially, thereby enhancing radial thermal conduction \citep{bal00,par05,par07}. Subsequent work has shown that even relatively gentle turbulence driven by galaxy motions and minor mergers can dominate these instabilities and arrange the magnetic field more isotropically \citep{par10, rus10, zuh13a}. Since in our simulations the turbulent velocity dispersion dominates over the Alfv\'en speed even at late times (Figure \ref{fig:sigma_vA}), we expect the effect of these instabilities to be swamped by the merger-driven turbulence, though simulations similar to ours including anisotropic thermal conduction would be required to verify this.
  
\section{Conclusions} \label{sec:conclusions}

In this work, we have analyzed a parameter space over mass ratio and impact parameter of idealized binary cluster mergers including the effects of magnetic fields. Our main conclusions are as follows:

\begin{itemize}

\item The bulk and turbulent flows created by merging clusters amplify magnetic fields in distinct ways. Compression of gas during core passages and behind shock fronts increases the magnetic energy in the core region overall. Shear flows created shortly after the first core passage of a merger along cold fronts and ram-pressure stripped gas amplify and stretch magnetic field lines, producing long, laminar magnetic structures which can stretch for $\sim$1-2~Mpc, which are easier to create and sustain in off-axis mergers. These structures are transient, however, and only last for a few Gyr at most, due to the turbulence driven by the second and following core passages. Otherwise, the velocities generated are turbulent in character, which generate turbulent magnetic fields. 

\item The magnetic energy of the clusters increases during the first and second core passages. At the first core passage, the dominant mechanism is the compression of magnetic field lines from the deepening of the gravitational potential and by shock fronts. For simulations with a non-zero impact parameter, shearing motions also increase the magnetic energy after this time. At later times, the dominant mechanism of amplification is stretching of magnetic fields by shear motions and turbulence. At the end of each simulation, as the merger remnant relaxes, the magnetic energy of the cluster gradually decreases. In general, simulations with smaller subclusters (increased mass ratio) generate less compressive motions and turbulence, and hence generate less amplification of the magnetic field, but simulations with larger impact parameters for a given mass ratio generate more magnetic energy via shearing motions.

\item At the final merged state of each simulation, the turbulent kinetic energy in the core region is $\sim$1-3~orders of magnitude higher than the magnetic energy, but only a factor of 10 or less higher outside the core region. The ratio between these quantities is largest in simulations with high mass ratio and impact parameter, since these cores have been less affected by strong turbulence and hence have less amplification of the magnetic field. 
 
\item The turbulent velocity dispersion of the gas in the cluster is very similiar in both merger simulations which include a magnetic field and those which do not. Due to this fairly insignificant effect of the magnetic field on the turbulent motions in the cluster, the final core entropies in each merger with magnetic fields included are nearly indistinguishable from those in the unmagnetized simulations. 
 
\item In the absence of radiative processes, the turbulent-driven mixing which increases the entropy of the core also decreases the density in the core region substantially. Since the magnetic field is frozen into the fluid, the average magnetic field strength also decreases substantially in the core region. Outside of the core region, the magnetic field of the final merger remnant is increased over the initial value. In the core region, the velocity and magnetic fields are very nearly isotropic, but outside of the cores these fields can become preferentially radial in on-axis mergers or preferentially tangential in off-axis mergers.
 
\end{itemize}

These results confirm in detail the fact stated simply at the outset: magnetic fields are generally not dynamically significant in the ICM of merging clusters of galaxies. The compressive and stretching actions on the magnetic field driven by the merger, though they do increase the magnetic field strength, do not increase it to the extent that it has a significant effect on the kinematic or thermodynamic properties of the merger remnant. This is in some contrast to the results of ZML11, who found that magnetic fields can have a dynamically significant effect on the cluster core in a relatively relaxed system with sloshing gas motions. However, the gas in those simulations was comparatively ``gently stirred'' by a gasless subcluster, and the typical turbulent velocities which resulted are much slower than encountered in these simulations, and thus comparable to the Alfv\'en velocity of the magnetic field. It is important to note that this conclusion holds {\it on average} in the bulk of the ICM--at regions where strong shear flows exist such as cold fronts, magnetic fields are still amplified to dynamically significant strengths ($\beta \sim$ 3-10, ZML11). 

Our binary merger simulations highlight the importance of gas compression, shear motions, and turbulence in amplifying the magnetic field during the most violent stages of cluster mergers. However, unlike what has been observed in cosmological simulations of cluster formation, during the period of relaxation after the formation of the merger remnant the magnetic energy decreases. This is likely due to a combination of the lack of radiative cooling in our simulations and the fact that real clusters undergo multiple mergers and are continuously stirred by substructure. Supporting this conclusion, \citet{sub06} argued that after a major merger the magnetic field would decay after the turbulence which drives its further growth had diminished, which is exactly what is seen in Figure \ref{fig:emag_evol}.

This investigation leaves room for further study. In the absence of a high angular-resolution X-ray observatory equipped with a microcalorimeter which can directly measure gas motions, a number of studies have used {\it Chandra} observations of surface brightness fluctuations in the ICM to indirectly probe the power spectrum of turbulent gas motions over a large range of scales \citep[e.g.][]{zhu15,chu16,zhu18}. Though this work shows that the effects of magnetic fields are not dynamically significant in the ICM during cluster mergers overall, it remains to be seen if localized field amplification on small scales can have an effect on the properties of observed X-ray surface brightness fluctuations. 

We did not include the effects of viscosity or thermal conduction in our simulations. As already mentioned, these processes will be highly anisotropic due to the wide separation in scales between the Larmor radii and mean free paths of the electrons and ions. The effects that these processes will have on the merger-driven gas motions may also have a non-negligible effect on the results described here. This is also left for future work. 

Finally, our non-radiative simulations neglect the important effects of radiative cooling, star formation, and stellar and AGN feedback. Without radiative cooling, the mixing of hot and cold gas produces a non-cool-core cluster as the final merger remnant, and the average magnetic field strength in its core region is lower than in the inital state, since the field is frozen in and the field lines become more spread apart as the gas becomes more dilute. In a real cluster with radiative cooling, the gas density would increase again and the field would become stronger. Future work will include the effects of cooling, star formation, and feedback in idealized cluster mergers to provide a more complete picture.

\acknowledgments
This work required the use and integration of a number of Python software packages for science, including Matplotlib \citep{hun07}\footnote{\url{http://matplotlib.org}}, NumPy \footnote{\url{http://www.numpy.org}}, SciPy \footnote{\url{http://www.scipy.org/scipylib/}}, and yt \citep{tur11}\footnote{\url{http://yt-project.org}}. We are thankful to the developers of these packages. The authors thank Paul Nulsen and Grant Tremblay for useful comments. JAZ acknowledges support through Chandra Award Number G04-15088X issued by the Chandra X-ray Center, which is operated by the Smithsonian Astrophysical Observatory for and on behalf of NASA under contract NAS8-03060. The numerical simulations were performed using the computational resources of the Advanced Supercomputing Division at NASA/Ames Research Center.

\appendix
\section{Supplemental Slice Plots}

\begin{figure*} [th!]
\begin{center}
\includegraphics[width=0.9\linewidth]{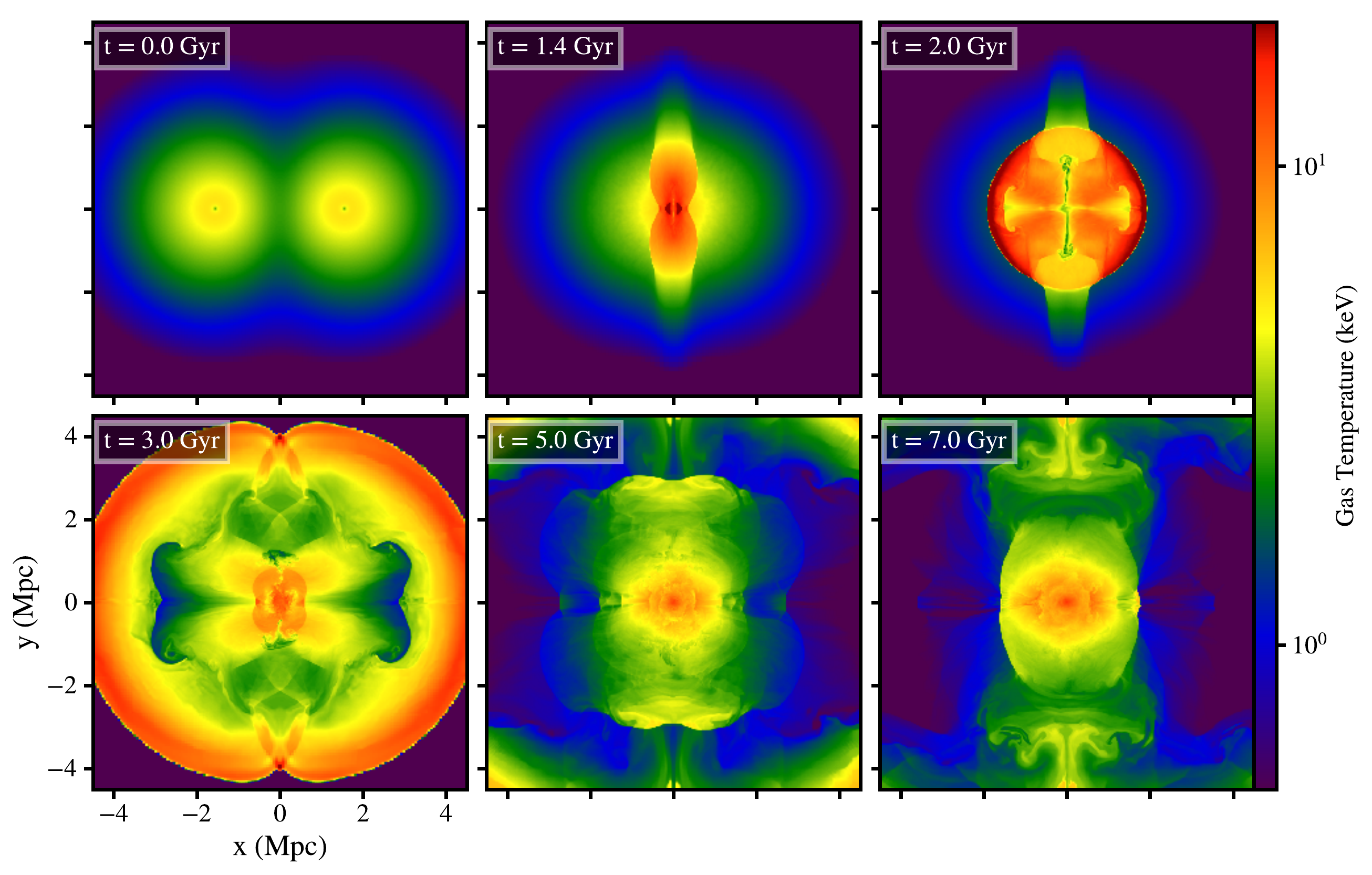}
\end{center}
\caption{Temperature slices through the collision axis for the MS1 simulation ($M_1/M_2 = 1$, $b = 0$). The epochs shown are: $t$ = 0, 1.4, 2.0, 3.0, 5.0, and 7.0 Gyr.\label{fig:1to1_b0_kT}}
\end{figure*}

\begin{figure*} [h!]
\begin{center}
\includegraphics[width=0.9\linewidth]{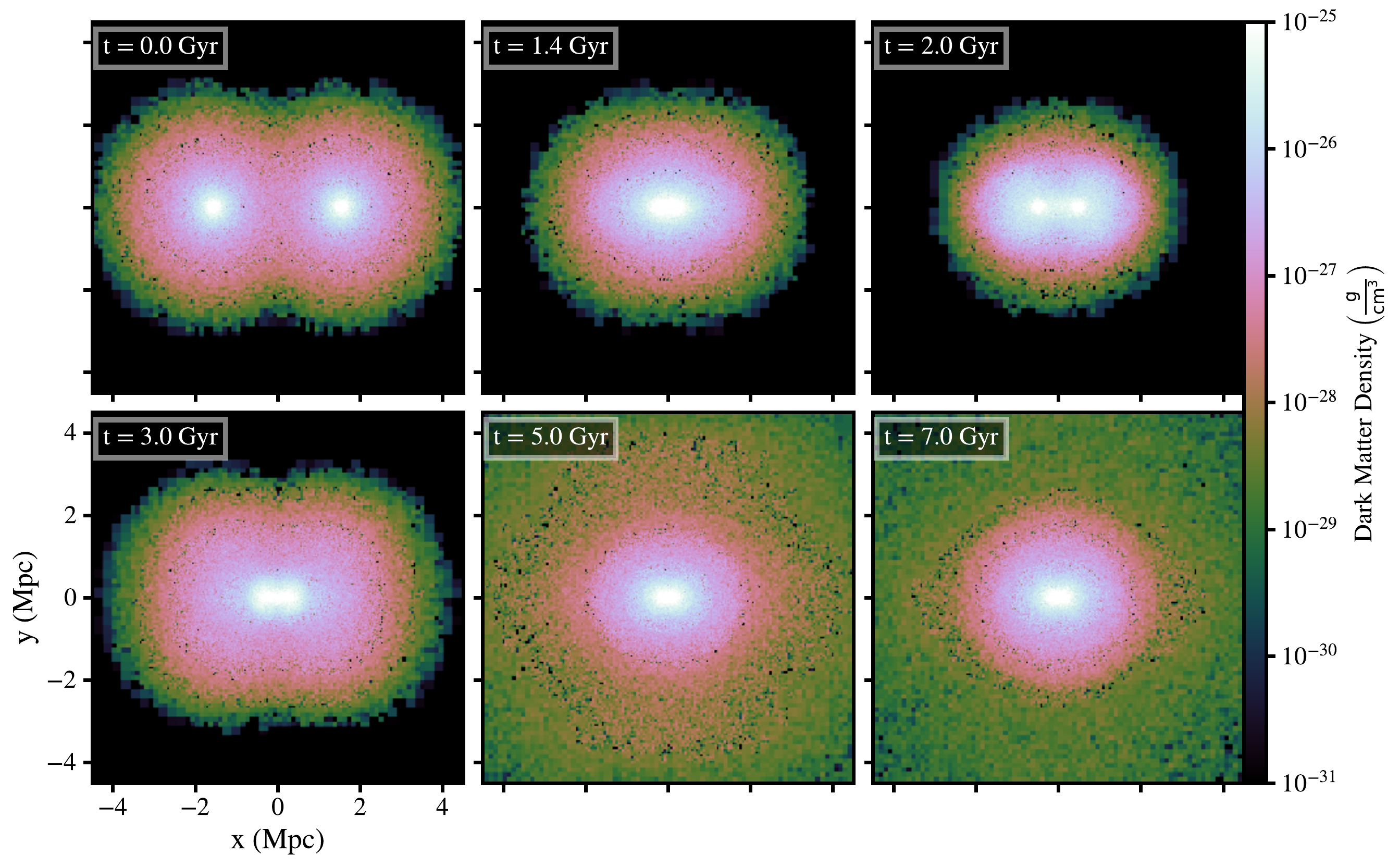}
\end{center}
\caption{DM density slices through the collision axis for the MS1 simulation ($M_1/M_2 = 1$, $b = 0$). The epochs shown are: $t$ = 0, 1.4, 2.0, 3.0, 5.0, and 7.0 Gyr.\label{fig:1to1_b0_all_cic}}
\end{figure*}

\begin{figure*} [t!]
\begin{center}
\includegraphics[width=0.9\linewidth]{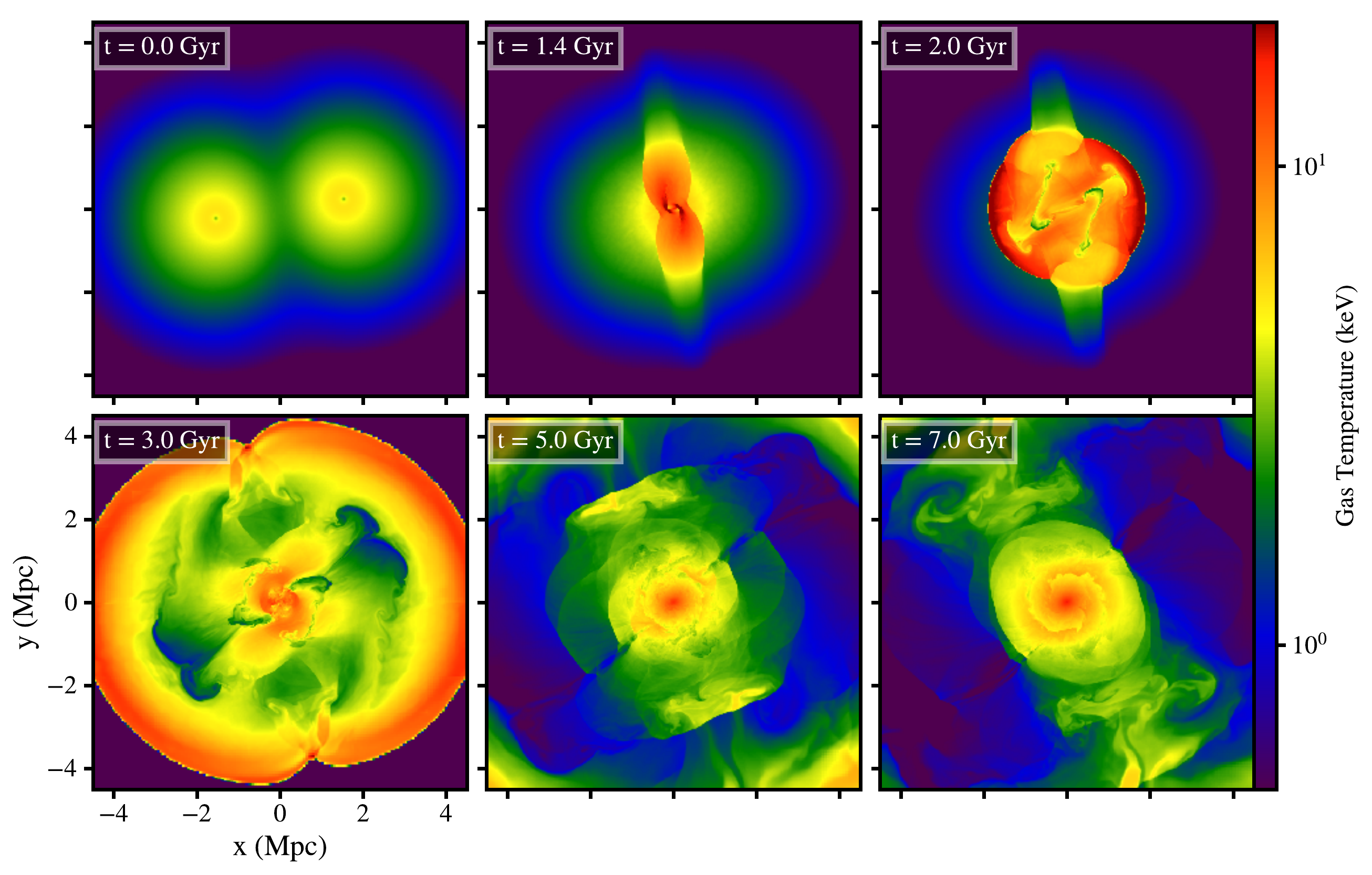}
\end{center}
\caption{Temperature slices through the collision axis for the MS2 simulation ($M_1/M_2 = 1$, $b = 0.3r_{200}$). The epochs shown are: $t$ = 0, 1.4, 2.0, 3.0, 5.0, and 7.0 Gyr.\label{fig:1to1_b0.5_kT}}
\end{figure*}

\begin{figure*} [t!]
\begin{center}
\includegraphics[width=0.9\linewidth]{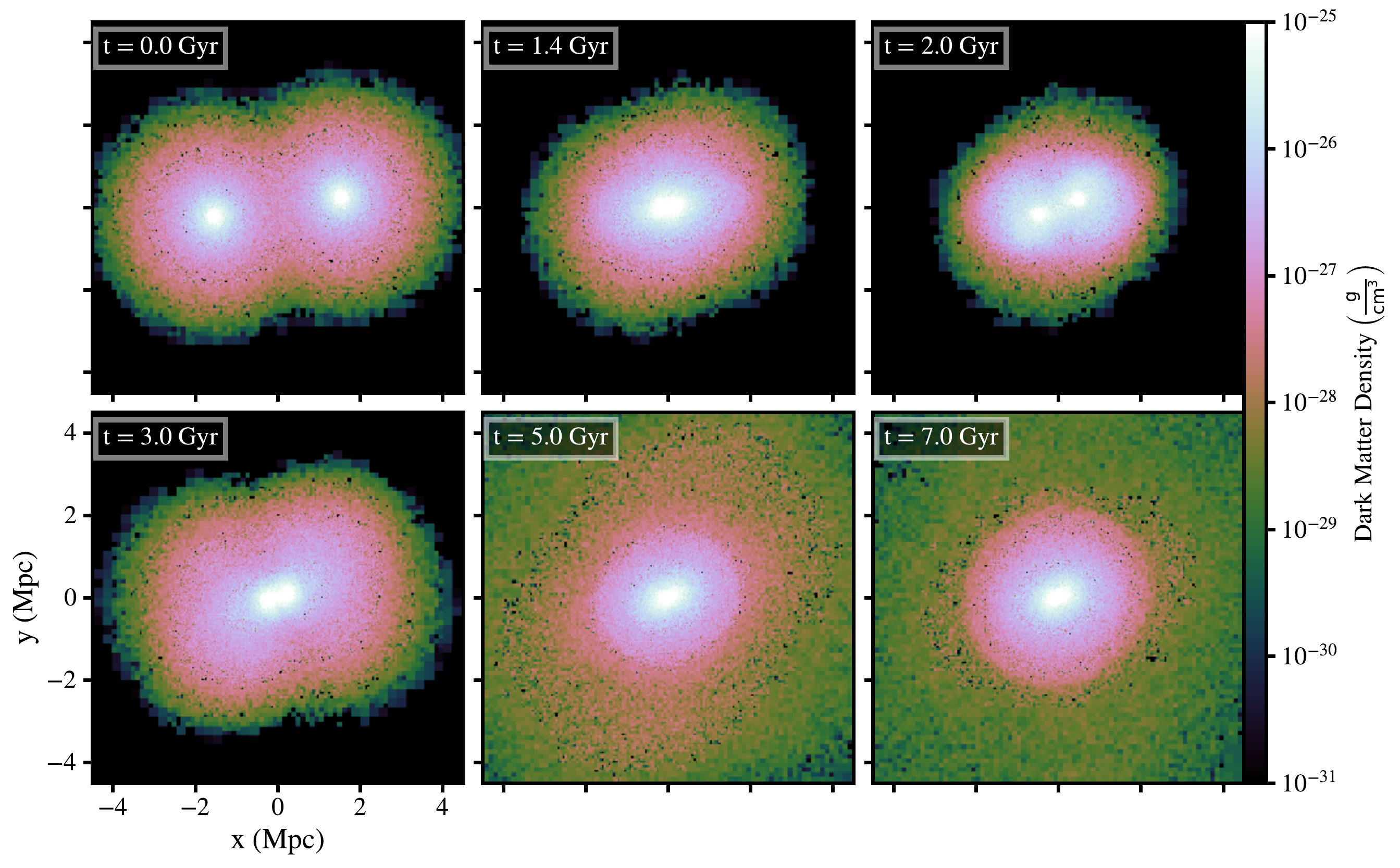}
\end{center}
\caption{DM density slices through the collision axis for the MS2 simulation ($M_1/M_2 = 1$, $b = 0.3r_{200}$). The epochs shown are: $t$ = 0, 1.4, 2.0, 3.0, 5.0, and 7.0 Gyr.\label{fig:1to1_b0.5_all_cic}}
\end{figure*}

\begin{figure*} [t!]
\begin{center}
\includegraphics[width=0.9\linewidth]{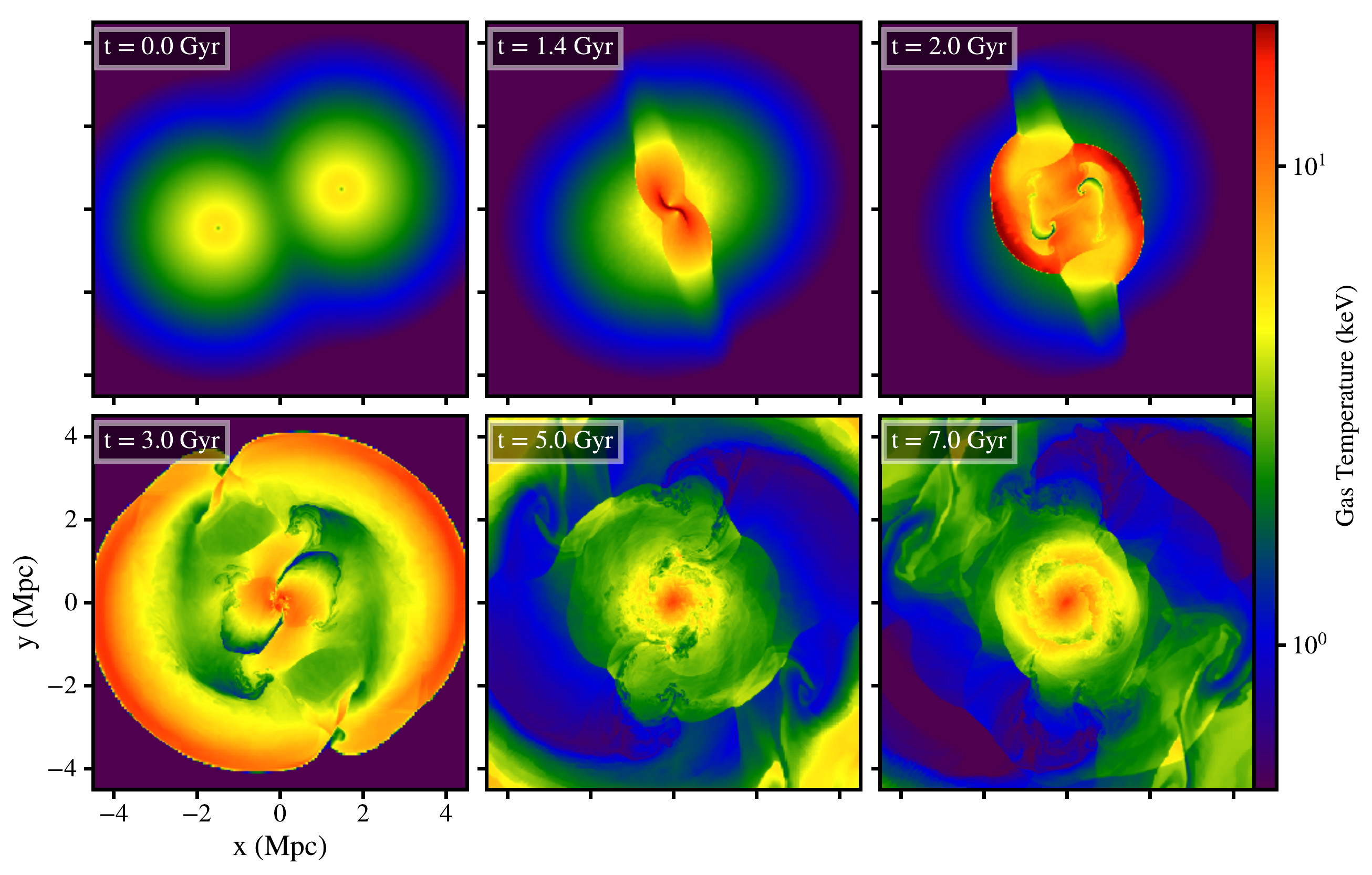}
\end{center}
\caption{Temperature slices through the collision axis for the MS3 simulation ($M_1/M_2 = 1$, $b = 0.6r_{200}$). The epochs shown are: $t$ = 0, 1.4, 2.0, 3.0, 5.0, and 7.0 Gyr.\label{fig:1to1_b1_kT}}
\end{figure*}

\begin{figure*} [t!]
\begin{center}
\includegraphics[width=0.9\linewidth]{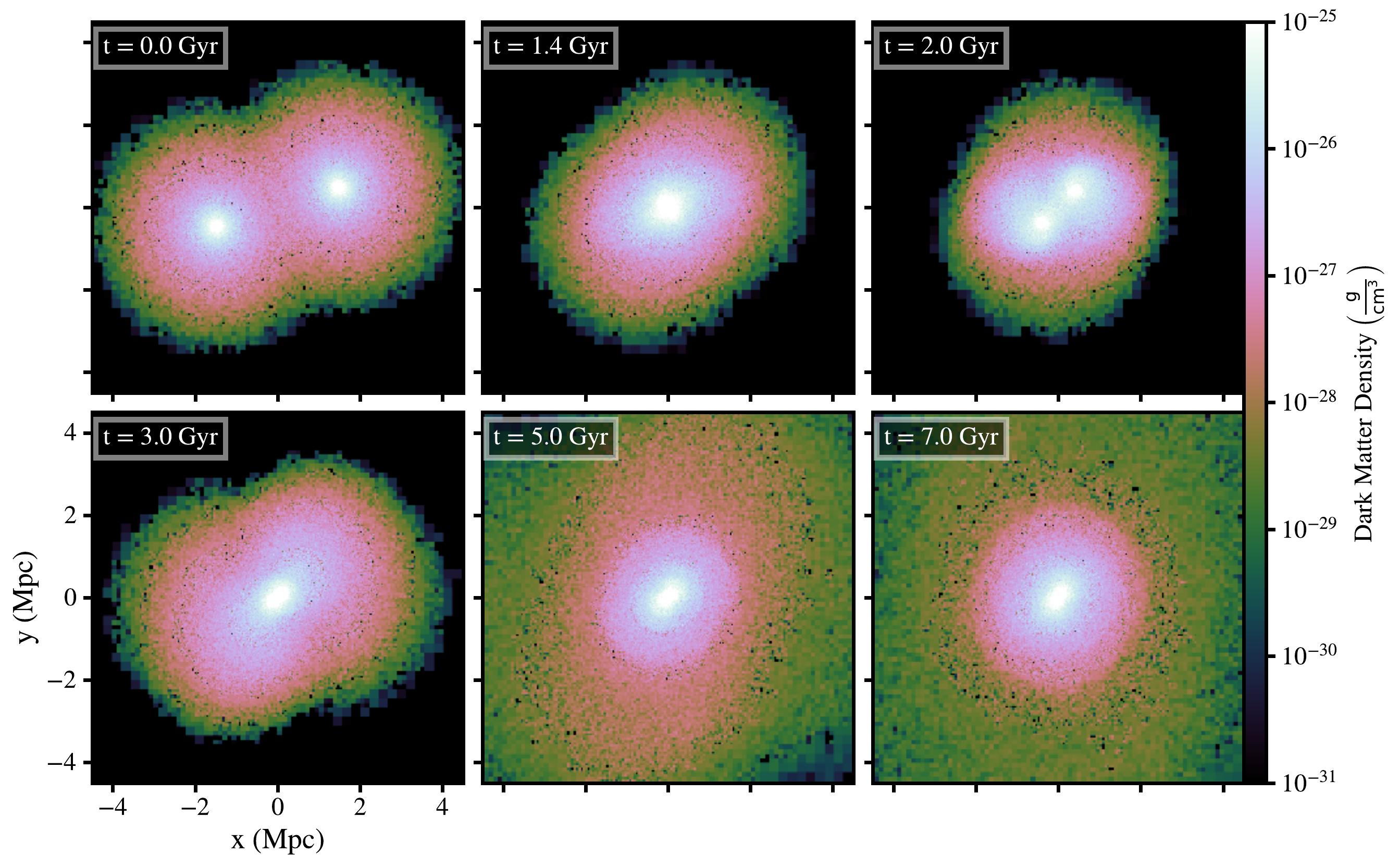}
\end{center}
\caption{DM density slices through the collision axis for the MS3 simulation ($M_1/M_2 = 1$, $b = 0.6r_{200}$). The epochs shown are: $t$ = 0, 1.4, 2.0, 3.0, 5.0, and 7.0 Gyr.\label{fig:1to1_b1_all_cic}}
\end{figure*}

\begin{figure*} [t!]
\begin{center}
\includegraphics[width=0.9\linewidth]{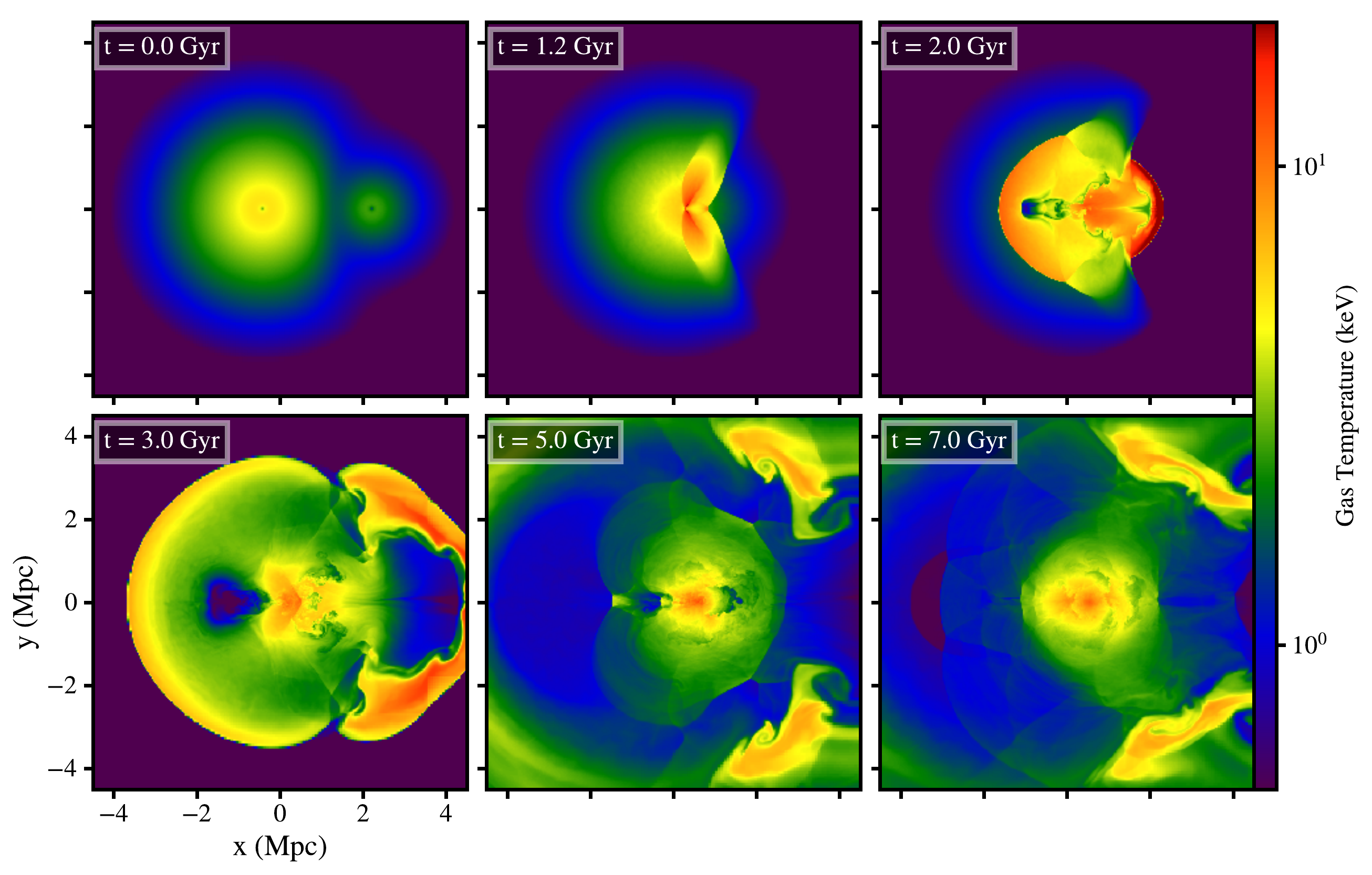}
\end{center}
\caption{Temperature slices through the collision axis for the MS4 simulation ($M_1/M_2 = 3$, $b = 0$). The epochs shown are: $t$ = 0, 1.2, 2.0, 3.0, 5.0, and 7.0 Gyr.\label{fig:1to3_b0_kT}}
\end{figure*}

\begin{figure*} [t!]
\begin{center}
\includegraphics[width=0.9\linewidth]{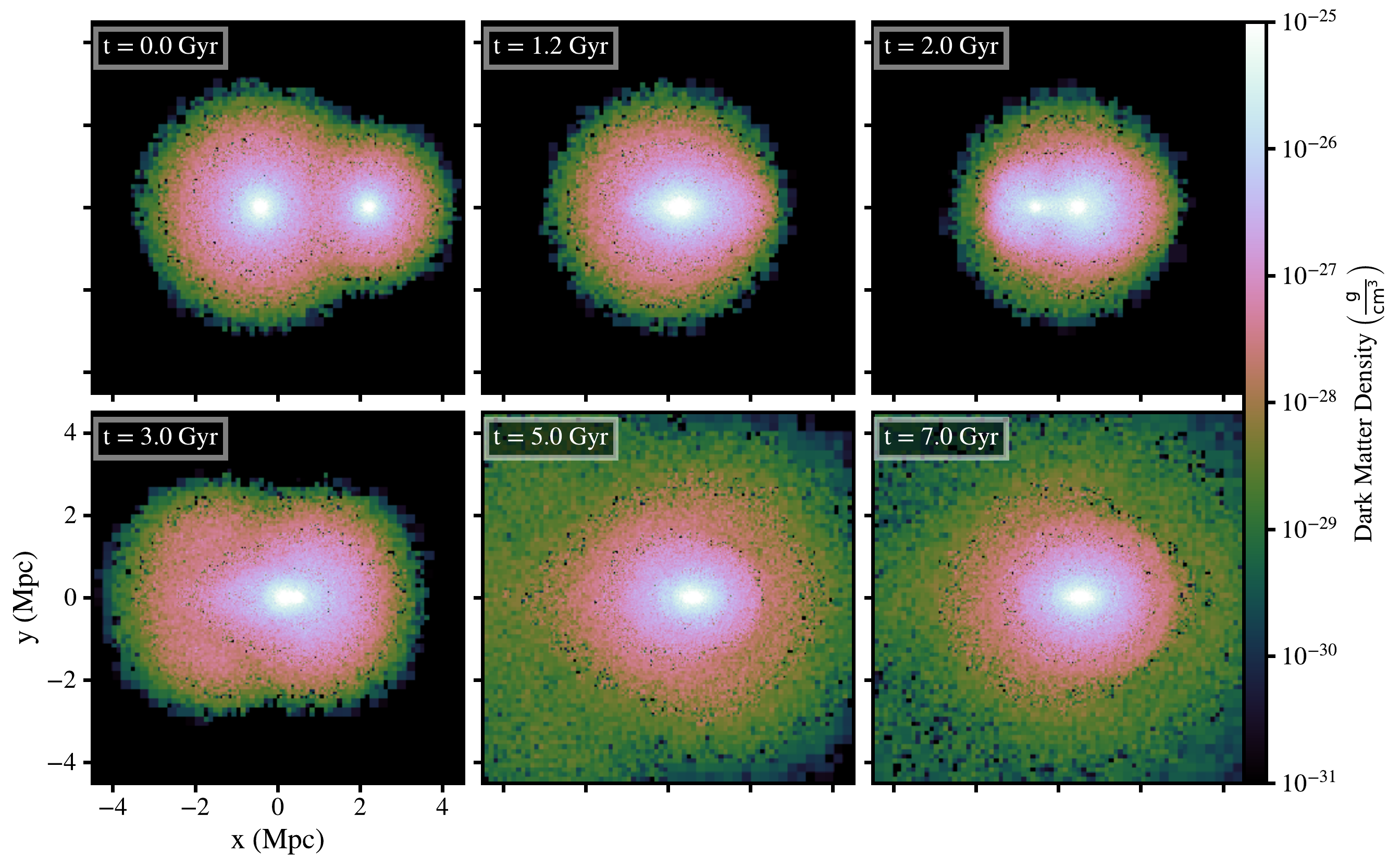}
\end{center}
\caption{DM density slices through the collision axis for the MS4 simulation ($M_1/M_2 = 3$, $b = 0$). The epochs shown are: $t$ = 0, 1.2, 2.0, 3.0, 5.0, and 7.0 Gyr.\label{fig:1to3_b0_all_cic}}
\end{figure*}

\begin{figure*} [t!]
\begin{center}
\includegraphics[width=0.9\linewidth]{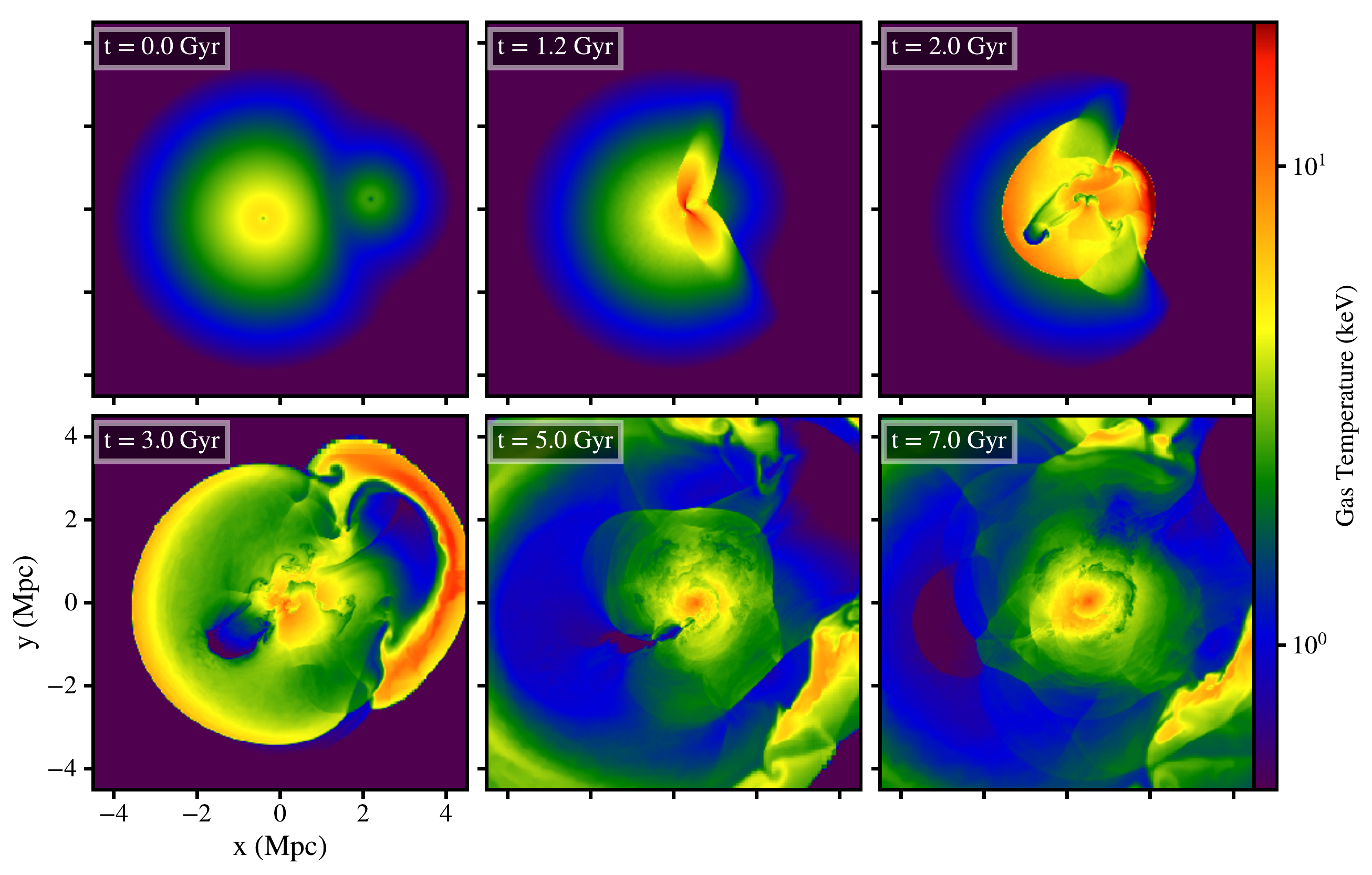}
\end{center}
\caption{Temperature slices through the collision axis for the MS5 simulation ($M_1/M_2 = 3$, $b = 0.3r_{200}$). The epochs shown are: $t$ = 0, 1.2, 2.0, 3.0, 5.0, and 7.0 Gyr.\label{fig:1to3_b0.5_kT}}
\end{figure*}

\begin{figure*} [t!]
\begin{center}
\includegraphics[width=0.9\linewidth]{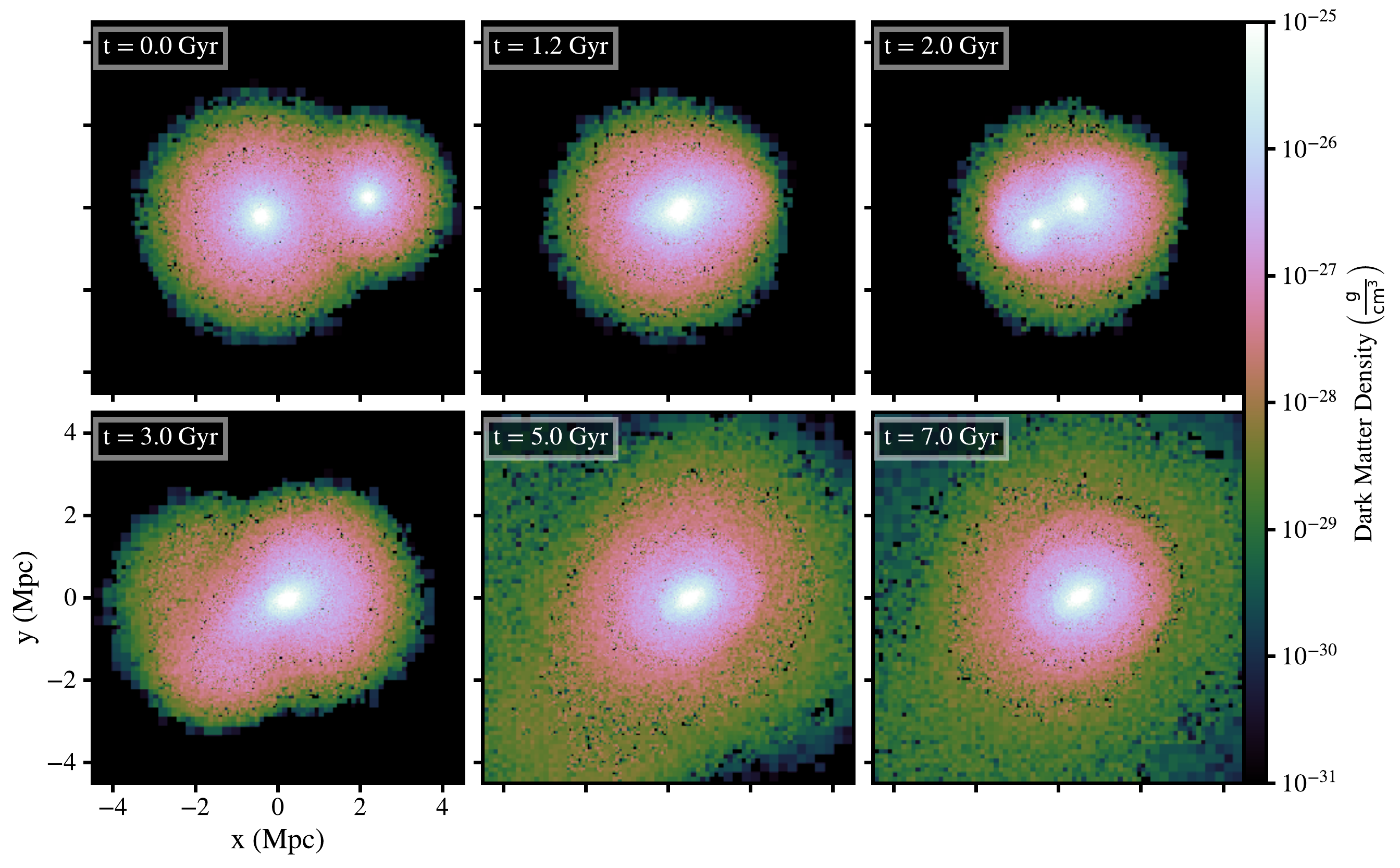}
\end{center}
\caption{DM density slices through the collision axis for the MS5 simulation ($M_1/M_2 = 3$, $b = 0.3r_{200}$). The epochs shown are: $t$ = 0, 1.2, 2.0, 3.0, 5.0, and 7.0 Gyr.\label{fig:1to3_b0.5_all_cic}}
\end{figure*}

\begin{figure*} [t!]
\begin{center}
\includegraphics[width=0.9\linewidth]{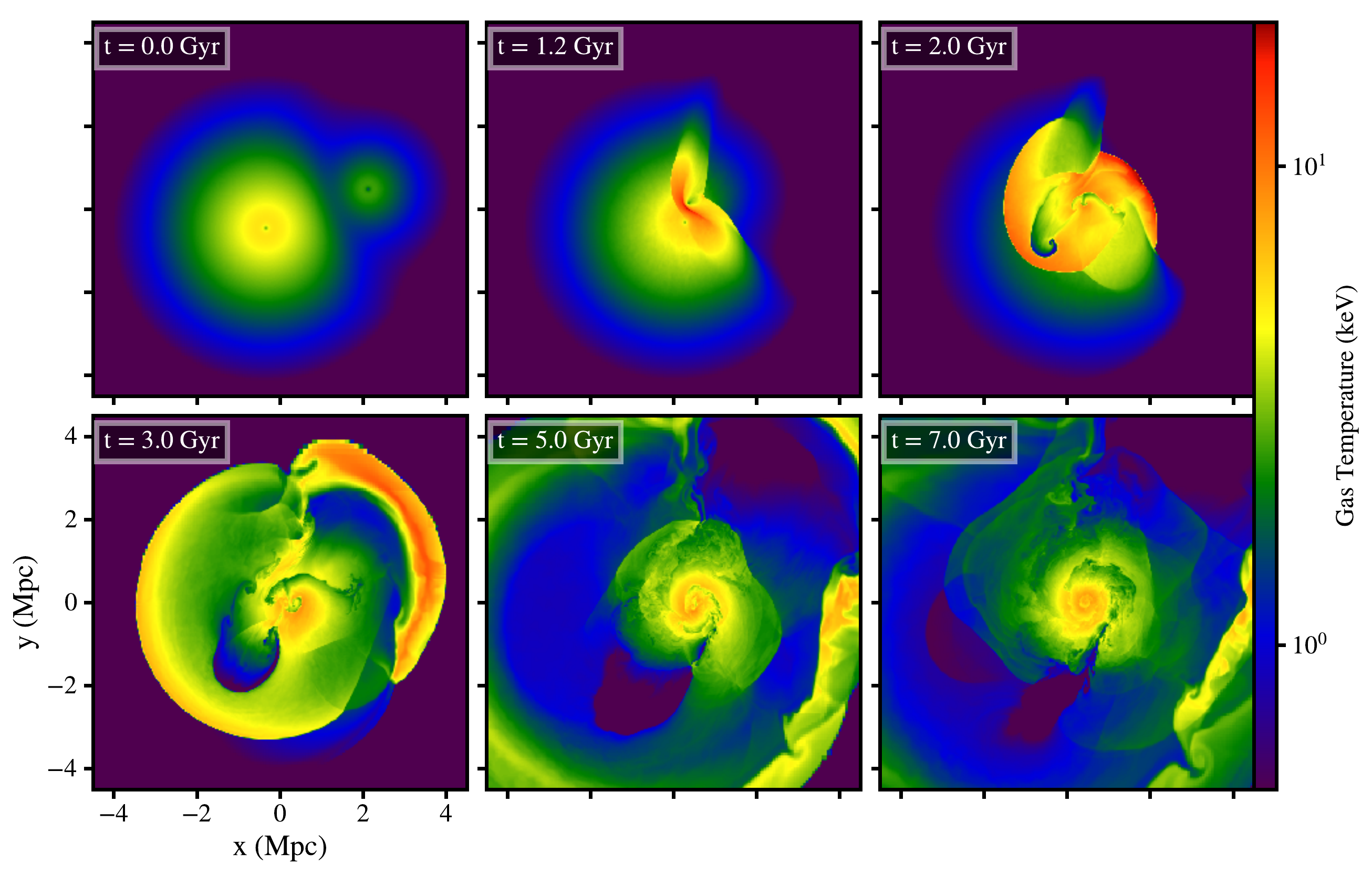}
\end{center}
\caption{Temperature slices through the collision axis for the MS6 simulation ($M_1/M_2 = 3$, $b = 0.6r_{200}$). The epochs shown are: $t$ = 0, 1.2, 2.0, 3.0, 5.0, and 7.0 Gyr.\label{fig:1to3_b1_kT}}
\end{figure*}

\begin{figure*} [t!]
\begin{center}
\includegraphics[width=0.9\linewidth]{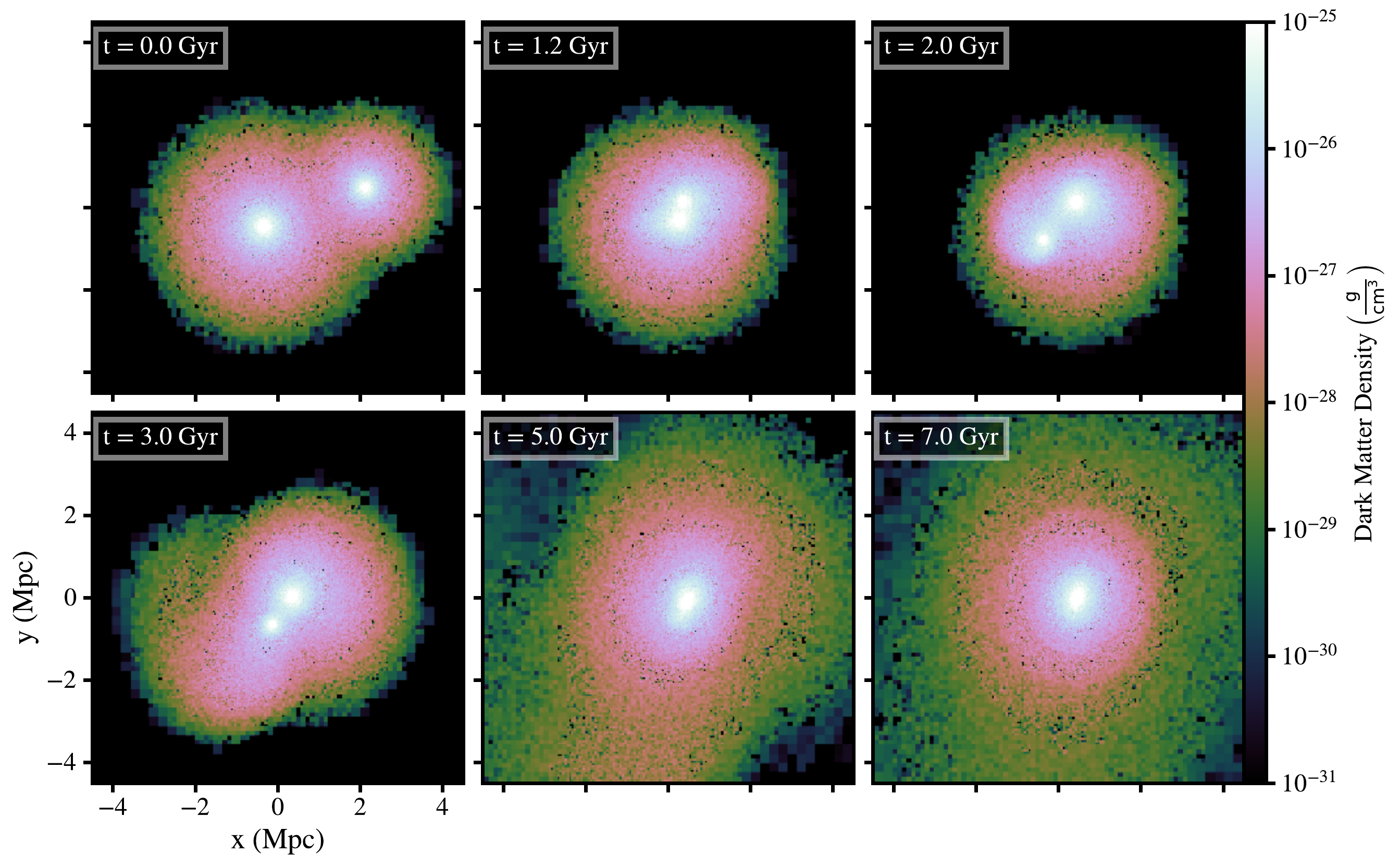}
\end{center}
\caption{DM density slices through the collision axis for the MS6 simulation ($M_1/M_2 = 3$, $b = 0.6r_{200}$). The epochs shown are: $t$ = 0, 1.2, 2.0, 3.0, 5.0, and 7.0 Gyr.\label{fig:1to3_b1_all_cic}}
\end{figure*}

\begin{figure*} [t!]
\begin{center}
\includegraphics[width=0.9\linewidth]{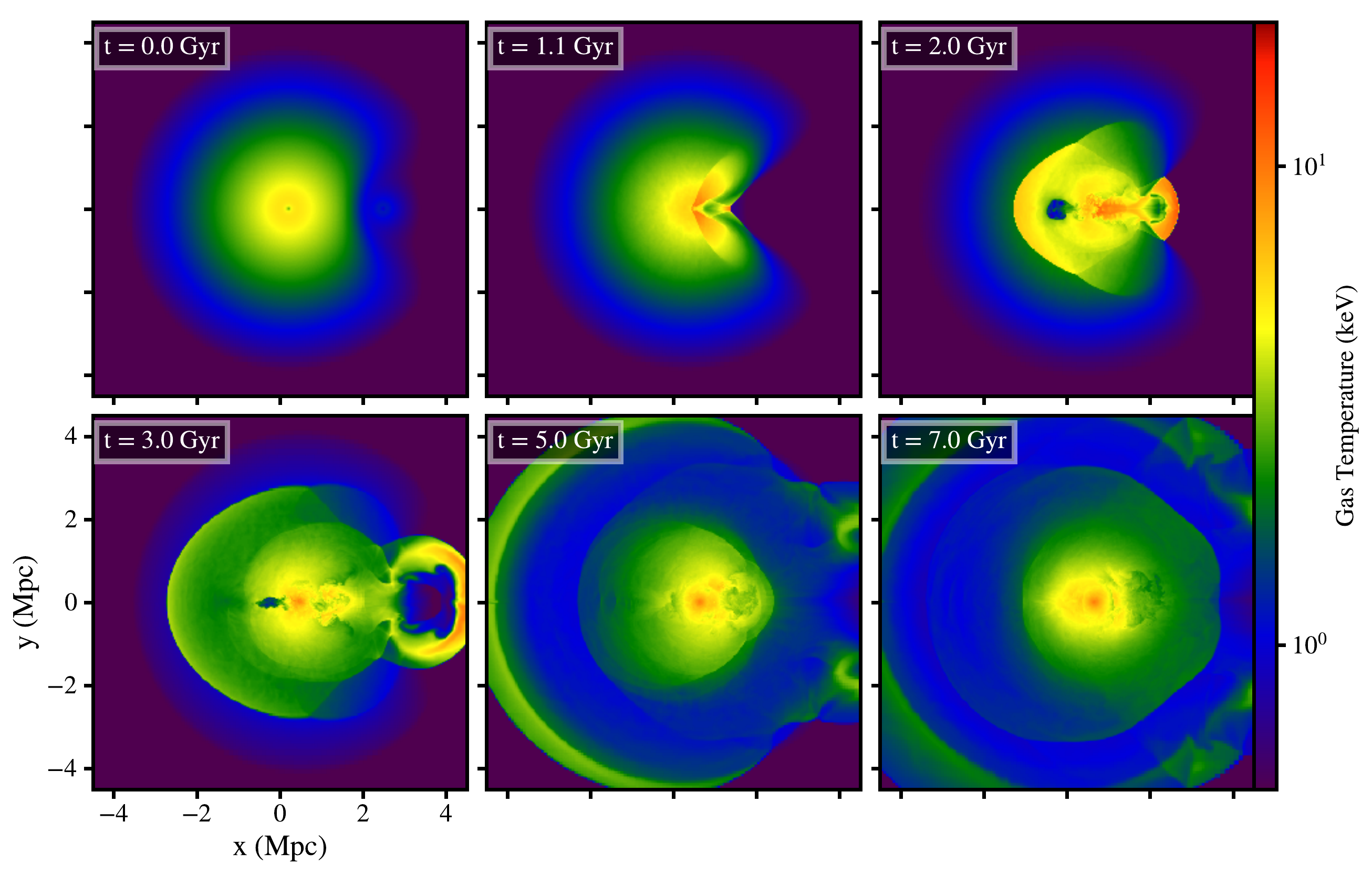}
\end{center}
\caption{Temperature slices through the collision axis for the MS7 simulation ($M_1/M_2 = 10$, $b = 0$). The epochs shown are: $t$ = 0, 1.1, 2.0, 3.0, 5.0, and 7.0 Gyr.\label{fig:1to10_b0_kT}}
\end{figure*}

\begin{figure*} [t!]
\begin{center}
\includegraphics[width=0.9\linewidth]{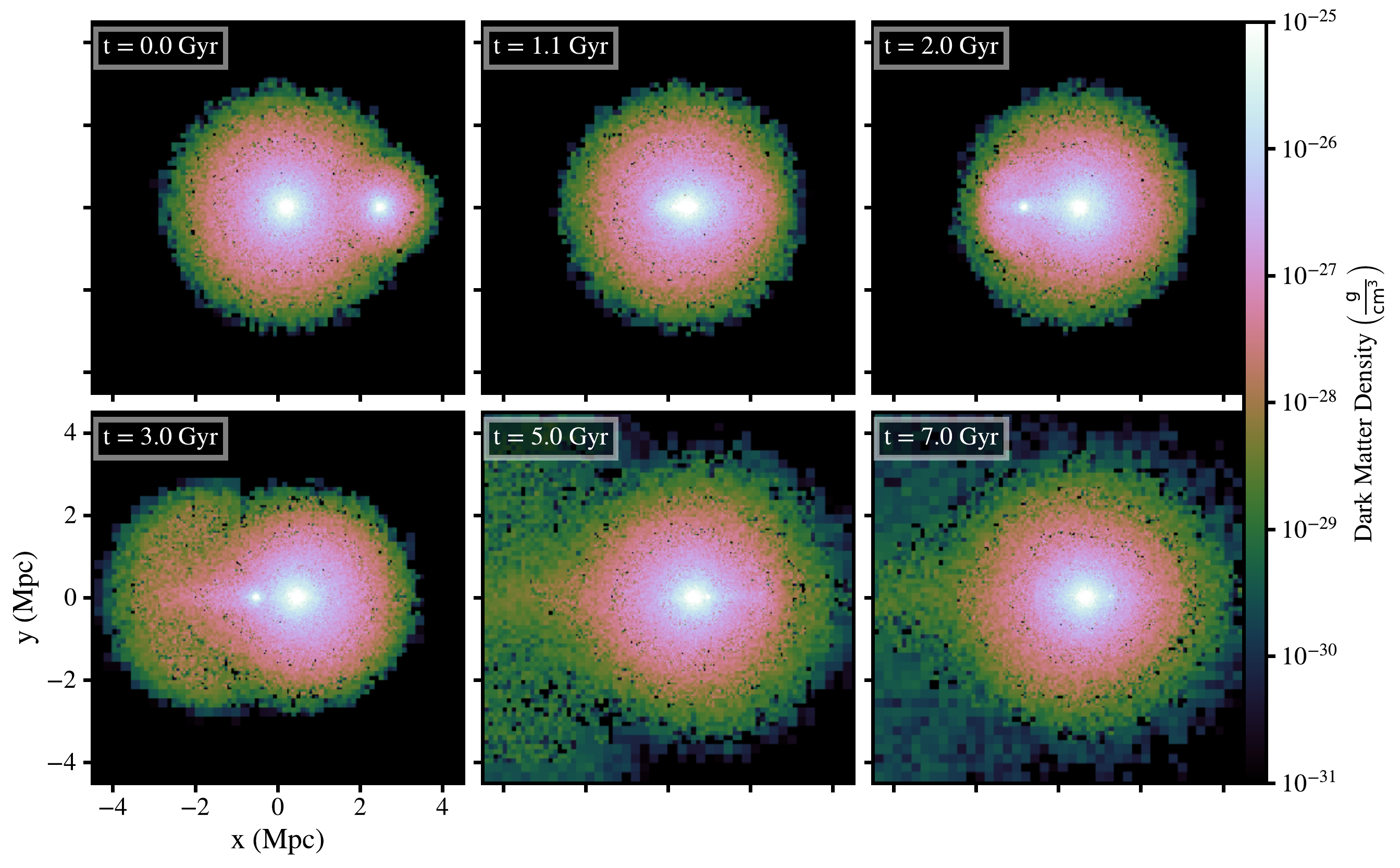}
\end{center}
\caption{DM density slices through the collision axis for the MS7 simulation ($M_1/M_2 = 10$, $b = 0$). The epochs shown are: $t$ = 0, 1.1, 2.0, 3.0, 5.0, and 7.0 Gyr.\label{fig:1to10_b0_all_cic}}
\end{figure*}

\begin{figure*} [t!]
\begin{center}
\includegraphics[width=0.9\linewidth]{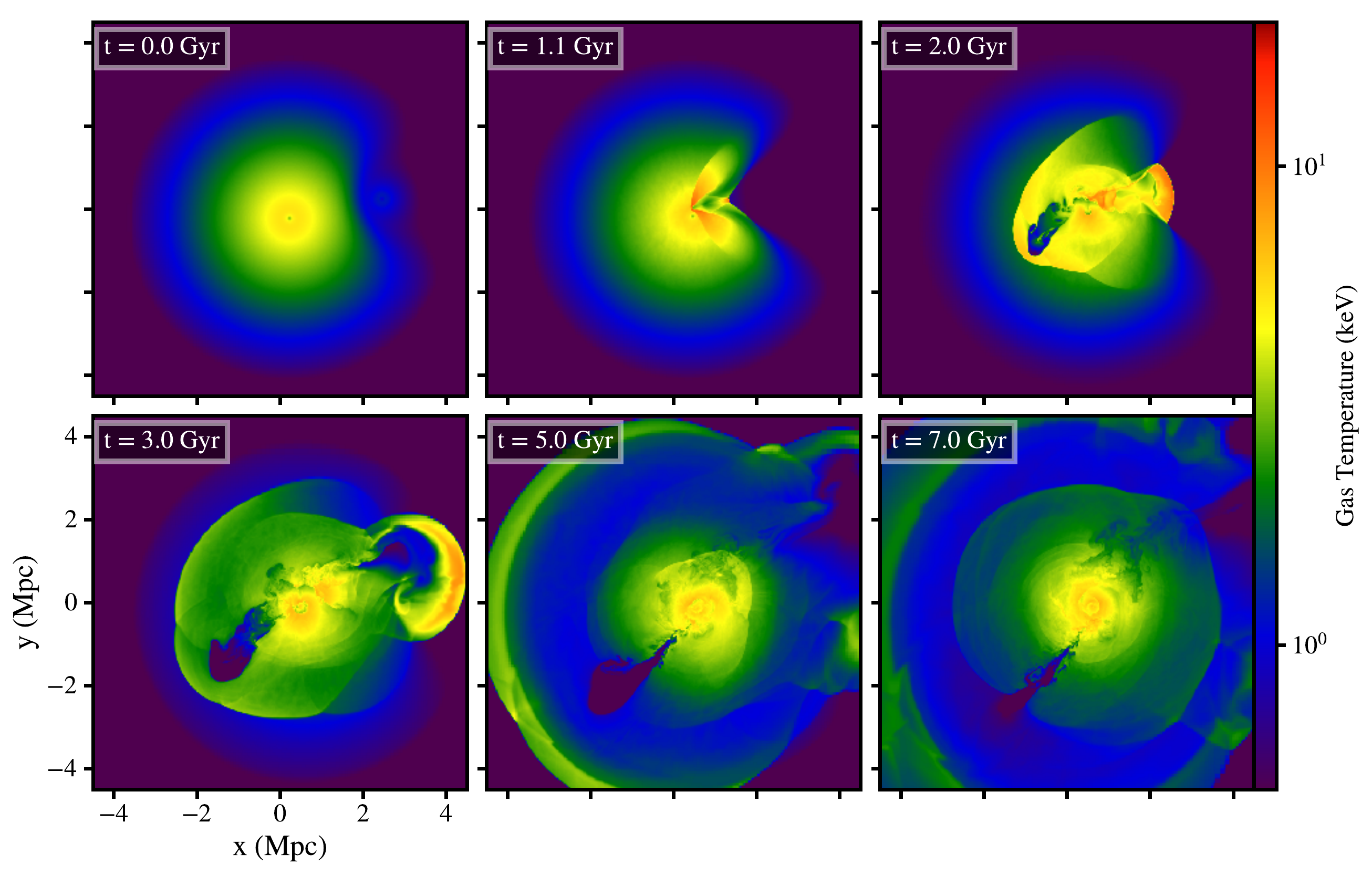}
\end{center}
\caption{Temperature slices through the collision axis for the MS8 simulation ($M_1/M_2 = 10$, $b = 0.3r_{200}$). The epochs shown are: $t$ = 0, 1.1, 2.0, 3.0, 5.0, and 7.0 Gyr.\label{fig:1to10_b0.5_kT}}
\end{figure*}

\begin{figure*} [t!]
\begin{center}
\includegraphics[width=0.9\linewidth]{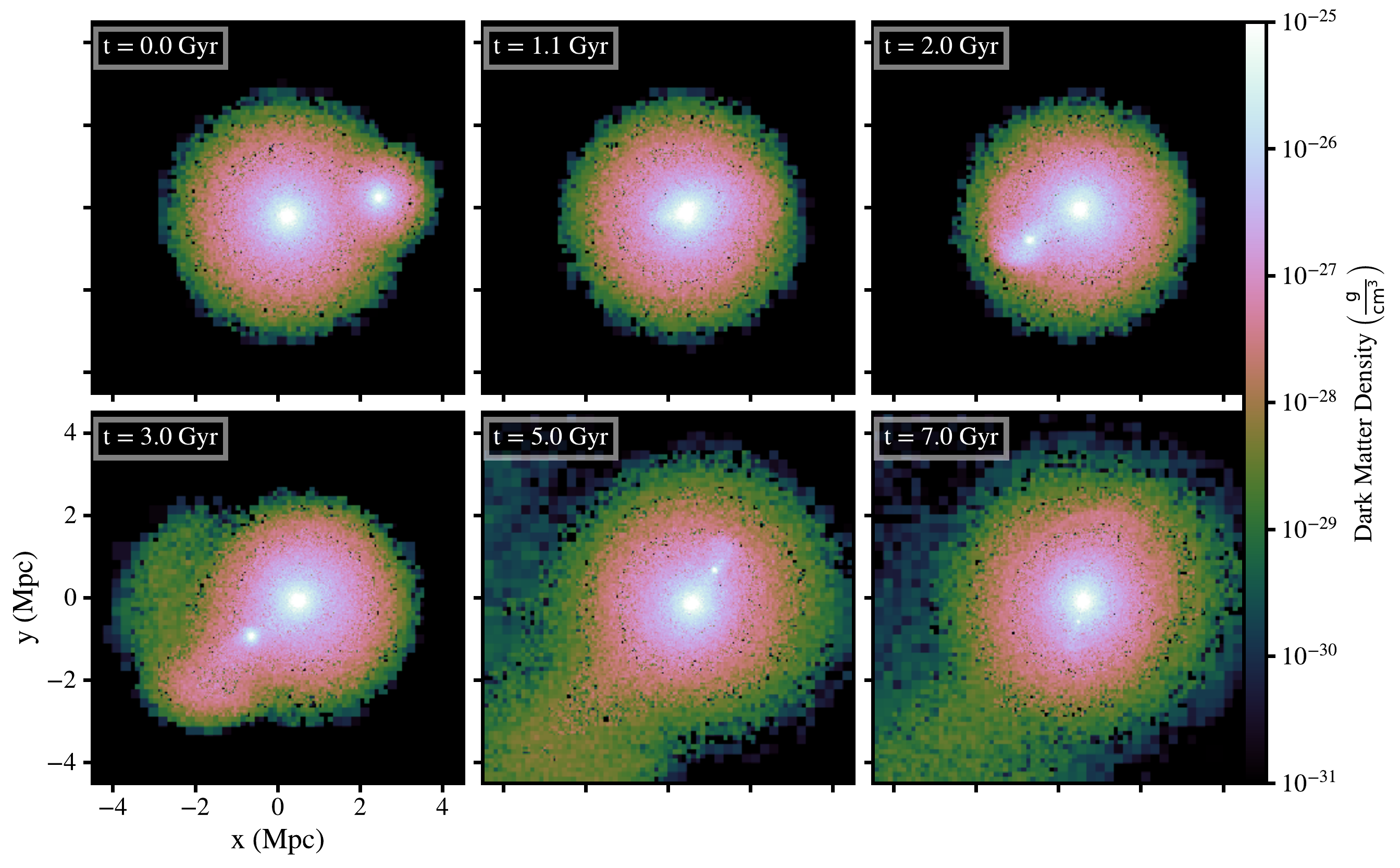}
\end{center}
\caption{DM density slices through the collision axis for the MS8 simulation ($M_1/M_2 = 10$, $b = 0.3r_{200}$). The epochs shown are: $t$ = 0, 1.1, 2.0, 3.0, 5.0, and 7.0 Gyr.\label{fig:1to10_b0.5_all_cic}}
\end{figure*}

\begin{figure*} [t!]
\begin{center}
\includegraphics[width=0.9\linewidth]{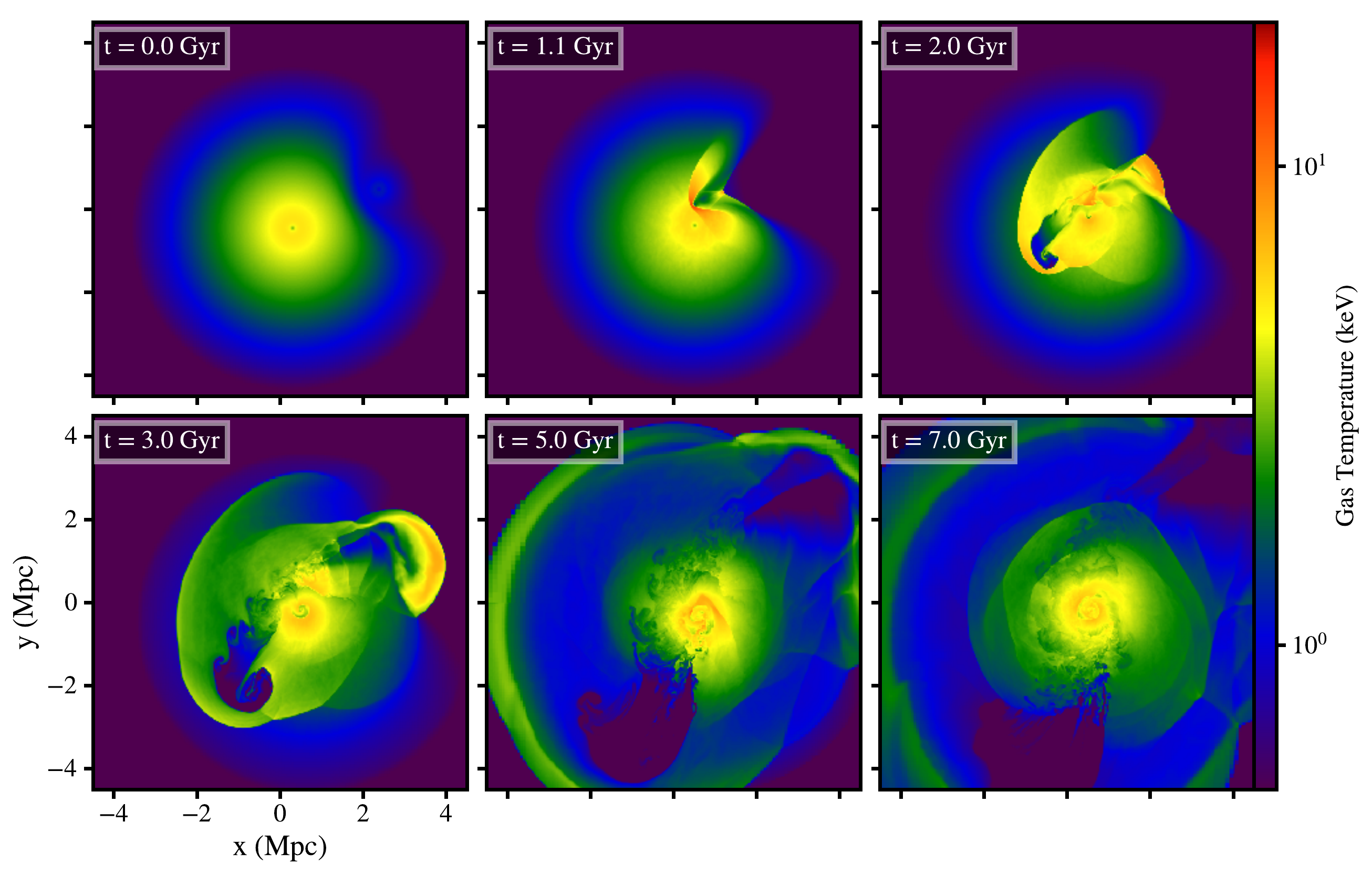}
\end{center}
\caption{Temperature slices through the collision axis for the MS9 simulation ($M_1/M_2 = 10$, $b = 0.6r_{200}$). The epochs shown are: $t$ = 0, 1.1, 2.0, 3.0, 5.0, and 7.0 Gyr.\label{fig:1to10_b1_kT}}
\end{figure*}

\begin{figure*} [t!]
\begin{center}
\includegraphics[width=0.9\linewidth]{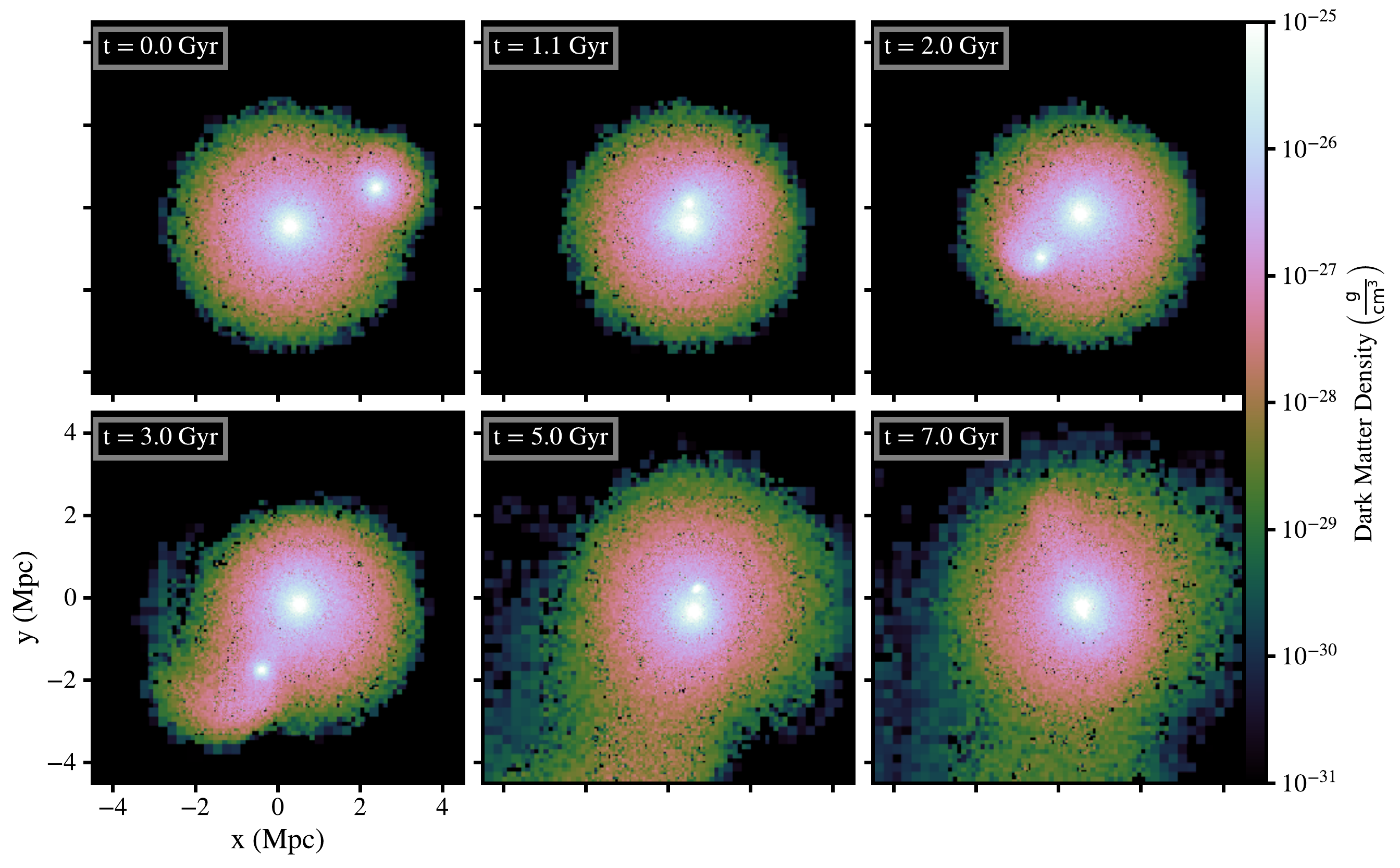}
\end{center}
\caption{DM density slices through the collision axis for the MS9 simulation ($M_1/M_2 = 10$, $b = 0.6r_{200}$). The epochs shown are: $t$ = 0, 1.1, 2.0, 3.0, 5.0, and 7.0 Gyr.\label{fig:1to10_b1_all_cic}}
\end{figure*}

We include additional slices of temperature and DM density in Figures \ref{fig:1to1_b0_kT}-\ref{fig:1to10_b1_all_cic}, which correspond to the same simulations and epochs as Figures \ref{fig:1to1_b0_density}-\ref{fig:1to10_b1_magfield}.


\end{document}